\documentclass[twocolumn]{aastex631}

\usepackage{lineno}
\usepackage{soulutf8}
\usepackage{ulem}
\usepackage[encoding,filenameencoding=utf8]{grffile}
\usepackage[caption=false]{subfig}

\def\ltsima{$\; \buildrel < \over \sim \;$}
\def\gtsima{$\; \buildrel > \over \sim \;$}
\def\lsim{\lower.5ex\hbox{\ltsima}}
\def\gsim{\lower.5ex\hbox{\gtsima}}
\usepackage{array}
\newcommand{\simu}[5]{N#1\_K#2\_RG#3\_TF#4\_W0#5}

\shorttitle{New parameters to trace the dynamical evolution of star clusters}
\shortauthors{Bhat et al.}

\begin{document}

\title{New Parameters for Star Cluster Dynamics: the role of clusters' initial conditions}

\author{B. Bhat}
\affiliation{Dept. of Physics and Astronomy ``A. Righi'', University
  of Bologna, Via Gobetti 93/2, Bologna, Italy}
\affiliation{INAF Osservatorio di Astrofisica e Scienza dello Spazio
  di Bologna, Via Gobetti 93/3, Bologna, Italy}

\author{B. Lanzoni}
\affiliation{Dept. of Physics and Astronomy ``A. Righi'', University
  of Bologna, Via Gobetti 93/2, Bologna, Italy}
\affiliation{INAF Osservatorio di Astrofisica e Scienza dello Spazio
  di Bologna, Via Gobetti 93/3, Bologna, Italy}

\author{E. Vesperini}
\affiliation{Dept. of Astronomy, Indiana University, Bloomington, IN 47401, USA}

\author{F. R. Ferraro}
\affiliation{Dept. of Physics and Astronomy ``A. Righi'', University
  of Bologna, Via Gobetti 93/2, Bologna, Italy}
\affiliation{INAF Osservatorio di Astrofisica e Scienza dello Spazio
  di Bologna, Via Gobetti 93/3, Bologna, Italy}

\author{F. I. Aros}
\affiliation{Dept. of Astronomy, Indiana University, Bloomington, IN 47401, USA}

\author{A. Askar}
\affiliation{Nicolaus Copernicus Astronomical Center, Polish Academy
  of Sciences, Bartycka 18, 00-716 Warsaw, Poland}

\author{A. Hypki}
\affiliation{Faculty of Mathematics and Computer Science,
  A. Mickiewicz University, Uniwersytetu Pozna\'nskiego 4, 61-614
  Pozna\'n, Poland}
\affiliation{Nicolaus Copernicus Astronomical Center, Polish Academy
  of Sciences, Bartycka 18, 00-716 Warsaw, Poland}

 
\begin{abstract}
We recently introduced three new parameters that describe the shape of
the normalized cumulative radial distribution (nCRD) of the innermost
stars in globular clusters and trace the clusters' dynamical
evolution. Here we extend our previous investigations to the case of a
large set of Monte Carlo simulations of globular clusters, started
from a broad range of initial conditions. All the models are analyzed
at the same age of 13 Gyr, when they have reached different
evolutionary phases. The sample of models is well representative of
the structural properties of the observed population of Galactic
globular clusters. We confirm that the three nCRD parameters are
powerful tools to distinguish systems in early stages of dynamical
evolution, from those that already experienced core collapse. They
might also help disentangle clusters hosting a low-mass intermediate-mass black hole of a few hundred solar masses, from cases with large concentrations of dark remnants in their centers. With respect to other dynamical indicators, the nCRD parameters offer the advantage of being fully empirical and easier to measure from observational data.
\end{abstract}



\section{Introduction} 
\label{sec:intro}
Globular clusters (GCs) are among the densest stellar environments in
the universe, comprising up to a few millions of stars in
approximately spherical configurations with typical values of the
half-mass radius of just a few parsecs.  Various internal dynamical
processes and the interaction with the external tidal field of the
host galaxy shape the present-day dynamical properties of GCs and
their stellar content.  Two-body relaxation, in particular, plays a
key role in driving GC evolution, and the manifestations of its
effects include segregation of massive stars in the cluster's inner
regions, and evolution towards energy equipartition.  One of the key
features of a cluster's internal dynamical evolution is the
progressive contraction of the core as the system approaches the core
collapse (CC) phase, which is eventually halted by the energy
generated by binary stars (see, e.g., \citealp{heggiehut2003}). If stellar mass black holes (BHs) are retained within the cluster, the BH sub-system experiences an early contraction just ~10-100 Myr after formation \citep{breen+13}.  This delays the CC of the observable stars until almost all BHs are ejected from the system. The cluster central density and central structure, and its post-CC
evolution depend on the fraction of primordial or dynamically formed
binaries, as well as the fraction of dark stellar remnants.  The
post-CC phase may be characterized by large fluctuations in the core
properties driven by gravothermal effects (see, e.g.,
\citealp{makino+1996} and references therein). The structural
evolution of a GC depends on the interplay among a number of factors
including, for example, the fraction of primordial binary stars, the
fraction of BHs retained in the cluster, the strength of
the external tidal field, and the cluster's mass, size, and internal
kinematics \citep[see, e.g.,][]{chernoff+90, mackey+07, Mocca_giersz,
  breen+2017, askar+18, kremer+2018, kremer+21, gieles+2021,
  pavlik+2022}. The high cluster densities, especially during the CC
and post-CC phases, may also enhance the frequency of close encounters
and physical collisions between stars and binaries inside the core,
making them efficient factories for the production of observable
astrophysical exotica, such as blue straggler stars, cataclysmic
variables, millisecond pulsars and tight binaries with compact
degenerate companions, including binary BHs that may lead to
gravitational wave events \citep[see, e.g.,][]{rodriguez+2015,
    rodriguez+2016, askar+2017, hong+2017, hong+2018, ye+20,
    antonini+2020}.  The degree of mass segregation and energy
  equipartition also strongly depends on the cluster's dynamical
  phase, its stellar content, and the fraction of dark remnants
  (e.g., stellar-mass and intermediate-mass BHs; see, e.g.,
  \citealp{Gill+08, alessandrini+16, peuten+2016, aros+23}).  The
  dynamical evolutionary stage of any observed GCs is thus influenced
  by the combination of various effects, and GCs with the same
  chronological ages typically have different dynamical ages, some
  having already reached CC, some other still being in early dynamical
  stages.

The observational identification of the dynamical phase of a GC and
the link between its dynamical history and the different properties
affecting its evolution can be quite challenging.  Typically, the
presence of a central power-law cusp in the observed density profile
of GCs in comparison with the flat core behavior of \cite{king66}
models is used as an indicator of CC or post-CC clusters
\citep{djorgovski+84, chernoff+89}. This diagnostic, however, can not
fully capture and identify the various possible dynamical phases of a
GC and the degree to which dynamical processes have altered the
cluster's internal dynamical properties and stellar content. For this
reason additional indicators have been proposed in recent years. These
indicators are either based on peculiar populations of heavy stars
(e.g., blue straggler stars) that are sensitive tracers of the
dynamical friction efficiency \citep{ferraro+12, ferraro+18,
  ferraro+19, ferraro+20, Ferraro+23, lanzoni+16}, or based on the
radial variations of the stellar mass function, kinematics, or
distribution of stars in the cluster \citep{baumgardt_makino03,
  webb_vesperini2017, tiongco+16, bianchini+2016, bianchini+2018,
  leveque+21, Bhat+22}.

In a recent study \citep[][hereafter B22]{Bhat+22} we defined three
new diagnostics of dynamical evolution based on the normalized
cumulative radial distribution (nCRD) of ``normal'' cluster stars,
i.e., stars observed at about the main-sequence turnoff level and
above.  The study has been performed by using Monte Carlo simulations
of star clusters, and it has been further extended in
\citet[][hereafter B23]{bhat+23} to investigate the effects of
different populations of primordial binary systems and stellar-mass
BHs. In those studies we found that the proposed new diagnostics can
efficiently distinguish clusters in various evolutionary phases,
including CC and post-CC, and we established a connection between them
and the various parameters (e.g., primordial binary fraction, content
of stellar-mass BHs) affecting the dynamics of GCs.  In those initial
investigations we have followed the time evolution of the new
diagnostics measured at different times for a few representative
systems.  In this paper we extend our investigation to explore the
evolution of GCs starting from a much broader range of initial
conditions. Instead of following an individual cluster at different
chronological ages, here we study the new parameters introduced in B22
in a sample of coeval clusters that, due to their different initial
conditions and evolutionary history, after 13 Gyr of evolution reached
different dynamical phases. The dataset we will focus our attention on
is thus similar to the observational population of Galactic GCs, which
have comparably old chronological ages (of $\sim 13$ Gyr), but
different dynamical ages.

The paper is organized as follows. In Section
\ref{sec:methods_initial_conditions_paper3}, we describe the initial
conditions of the Monte Carlo simulation survey and the methodology
adopted in the following analysis. In Section \ref{sec:analysis}, we
discuss the time evolution of the 1\% Lagrangian radius of the
extracted snapshots, we show the comparison between the structural
parameters of the simulated clusters and those observed in the
Galactic GC population, we presents the nCRDs of survey simulations,
and recall the definition of the three dynamical indicators.  In
Section \ref{sec:results} we describe the reference boundaries and
models adopted for the analysis of the nCRD parameters, elaborate on
the impact of the initial conditions, and discuss the ability of new
diagnostics to distinguish star clusters on the basis of their
dynamical evolutionary stage. Finally, the summary and conclusion of
the work are provided in Section \ref{sec:results}.


\section{Methods and Initial Conditions}
\label{sec:methods_initial_conditions_paper3}
The GCs analyzed in this work have been simulated through the MOCCA
Monte Carlo code which includes the effects of two-body relaxation, a
tidal truncation, binary-single and binary-binary encounters and, following the prescriptions based on the SSE and BSE
  codes by \citet{hurley+00, hurley+02}, also the effects of
binary and single stellar evolution \citep{Hypki_Giersz_2013,
  Mocca_giersz}. Massive star winds and BH masses are obtained from the standard SSE/BSE prescriptions. The wind mass loss prescriptions on the MS are based on \cite{nieuwenhuijzen+90} and \cite{Humphreys+94} for mass loss in luminous blue variable stars. We run a total of 108 simulations started from different combinations of initial conditions for the density profile
(for which we adopted \citealp{king66} models with different values of
the central dimensionless potential $W_0$), number of stars $N$
(defined as the sum of the number of single and binary stars,
$N=N_s+N_b$), tidal filling factor (TF, defined as the ratio of the
half-mass to the tidal radius, $r_h/r_t$), and galactocentric distance
$R_g$.  In all cases, the metallicity is $Z=10^{-3}$,
and the primordial binary fraction, defined as the ratio
$N_b/(N_s+N_b)$, is equal to 10\%.  As in B22, the binary properties
are set according to the eigenevolution procedure outlined in
\citet{kroupa95} and \citet{kroupa+2013}.  The stellar masses range
between 0.1 and $100 M_\odot$ following a \citet{IMF} initial mass
function. Supernova natal kick velocities for BHs are assigned either
according to a Maxwellian distribution with velocity dispersion of 265
km s$^{-1}$ \citep{Hobbs_05}, or according to the mass fallback
procedure described by \citet{belczynski+02}, where BHs have a reduced
kick velocity, hence a higher retention fraction \citep{arcasedda+18,
  askar+18}. We use $k=0$ and $k=1$ to flag models adopting kick velocities following the Maxwellian distribution from \cite{Hobbs_05} and with reduced kick velocity for BHs,
respectively. Remnant BH masses range from 5$M_\odot$ to up to 20$M_\odot$ (which is maximum BH mass from single star evolution at Z=0.001 in standard SSE). For models in which mass fallback prescriptions from \cite{belczynski+02} are enabled, the maximum BH mass from single stellar evolution is about $\sim$30$M_\odot$. After 50 Myr the fraction of BHs retained in the clusters in simulations with fallback on is about 45-50\%. The initial values adopted for the different parameters
are summarized in Table \ref{tab:list of sims_paper3}. Their
combination corresponds to a total of 108 runs. To identify each of
these simulations, we introduce a naming convention based on their
initial conditions. For instance, \simu{500}{0}{2}{0.1}{5} represents
the simulation run with N=500K, $k=0$, $R_g$=2 kpc, TF=0.1 and
$W_0$=5.  Out of 108 simulated GCs, only 92 survived after 13 Gyr of
evolution. As in the previous studies (B22, B23), they have been
analyzed from the point of view of an observer, meaning that standard
procedures and approximations adopted in observational works have been
applied: in each snapshot, the simulated cluster is projected onto a
2D plane, and a distance of 10 kpc from the observer has been assumed
to transform the distances from the centre of the system from parsecs
to arcseconds. In addition the magnitude of binary systems is computed
by summing up the luminosities of both components in analogy to what
is done in observed GCs where they can not be resolved.

\begin{table}
 \caption{Initial conditions of the simulations}
 \label{tab:list of sims_paper3}
\begin{center}
    \bigskip
    \begin{tabular}{l l l l}
    \hline
    \hline
      $N$   & 500K & 750K &  1M\\
      \hline
      $W_0$ & 5 & 7 \\
      \hline
      $k$   & 0 & 1\\
      \hline
      $R_g$ [kpc]   & 2 & 4 & 6\\
      \hline
      TF ($r_h/r_t$)  & 0.025 & 0.05 & 0.1 \\
      \hline
      \hline
    \end{tabular}
\end{center}
Values of the number of stars ($N$), King dimensionless central
potential ($W_0$), BH kick velocity parameter ($k$), galactocentric
distance ($R_g$), and tidal filling factor (TF=$r_h/r_t$) adopted as
initial conditions in the Monte Carlo simulations analyzed in this
work. The simulations have been run with all the
  possible combinations of the parameters listed in the table,
  resulting in a total of 108 models.
\end{table}


\section{Analysis of the simulated clusters}
\label{sec:analysis}

\begin{figure}[ht]
    \centering
    \includegraphics[width=0.5\textwidth]{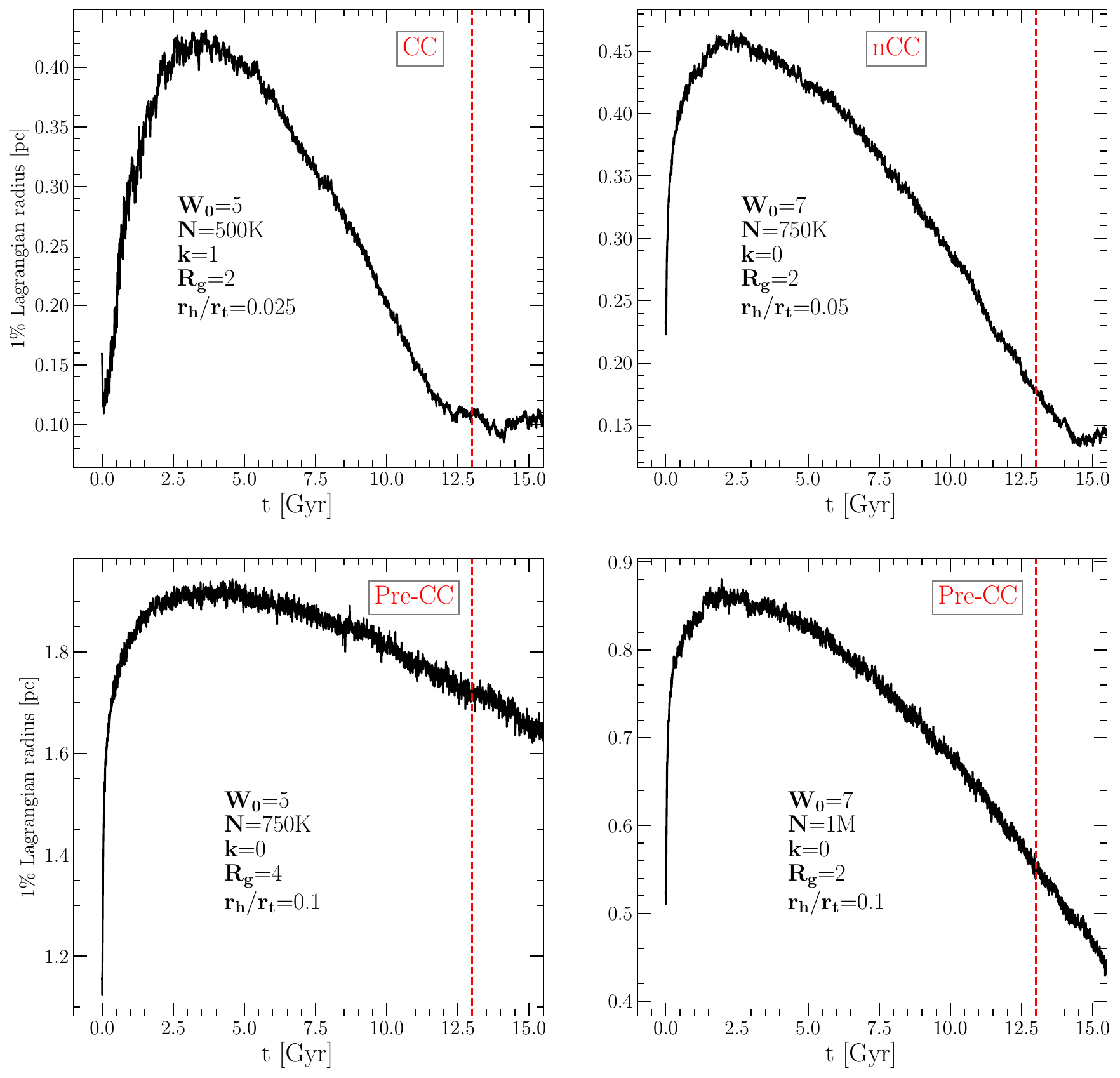}
    \caption{Time evolution of the 1\% Lagrangian radius for the
      simulation runs \simu{500}{1}{2}{0.025}{5},
      \simu{750}{0}{2}{0.05}{7}, \simu{750}{0}{4}{0.1}{5} and
      \simu{1}{0}{2}{0.1}{7} (from top left, to bottom right). The red
      vertical line at 13 Gyr marks the time snapshot of the
      simulation used for analysis. The classification of the 13 Gyr
      time snapshots in terms of dynamical evolutionary stage (CC,
      near to CC, and pre-CC) is marked in each panel.}
    \label{fig:lr1_cc_nocc}
\end{figure}

\begin{figure*}[ht]
    \centering
    \subfloat
    {\includegraphics[width=0.495\textwidth]{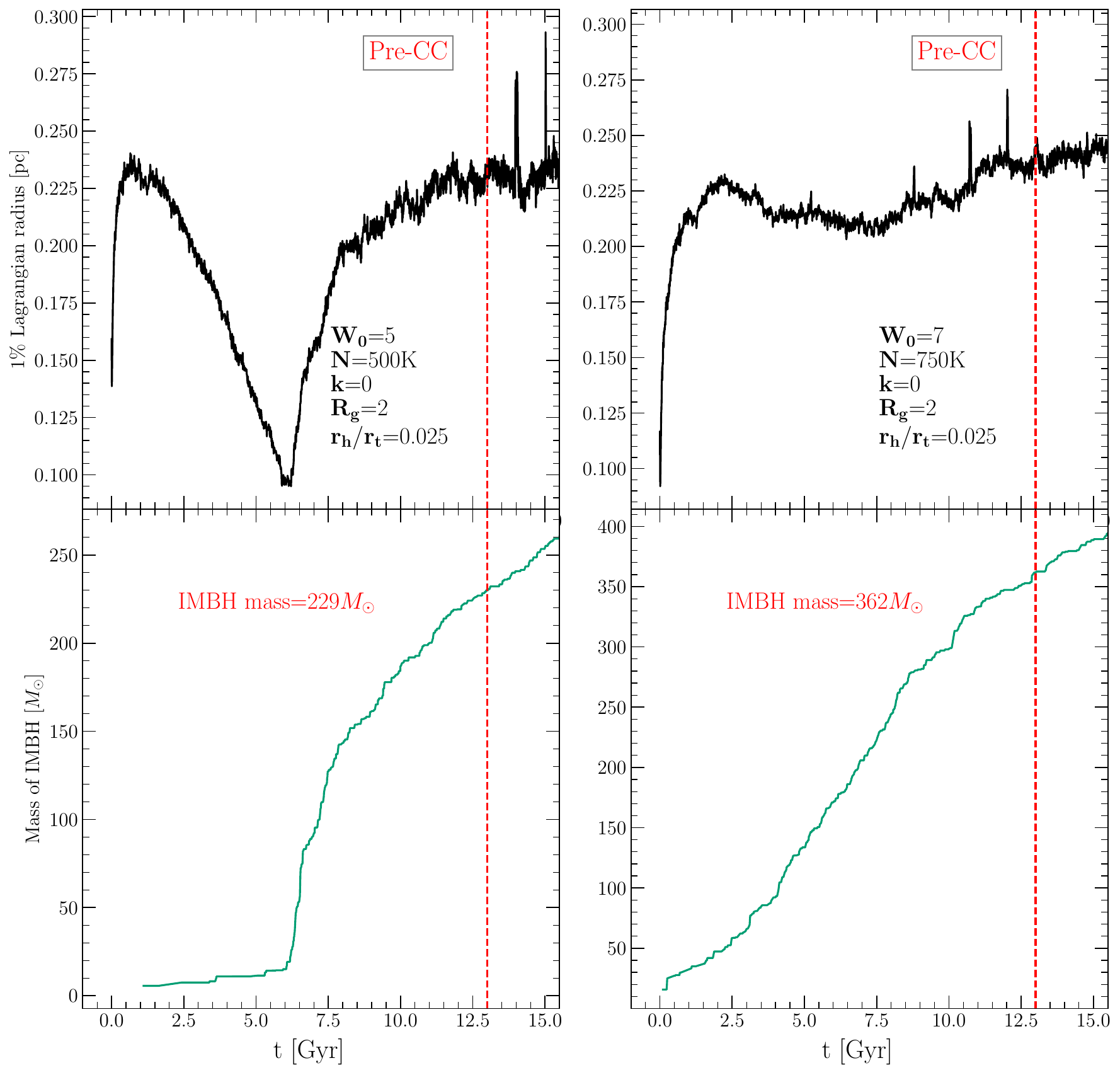}}
    \hfill
    \subfloat
    {\includegraphics[width=0.49\textwidth]{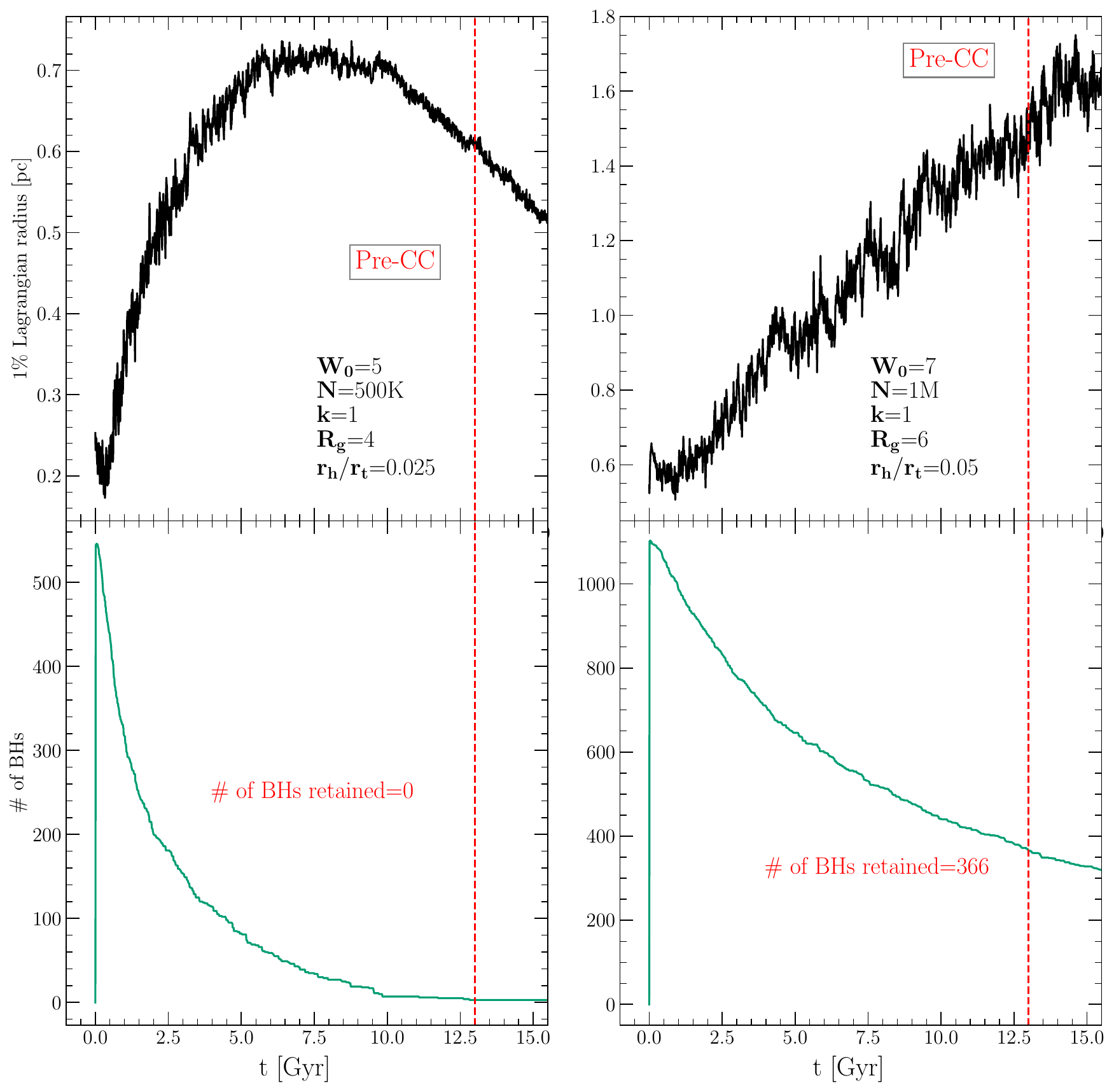}}
    \hfill
    \caption{Top row: as in Fig.\ref{fig:lr1_cc_nocc}, but for the
      simulations \simu{500}{0}{2}{0.025}{5},
      \simu{750}{0}{2}{0.025}{7}, \simu{500}{1}{4}{0.025}{5} and
      \simu{750}{1}{6}{0.05}{7} (from left to right). Bottom row: time
      evolution of the IMBH mass (first and second columns) and of the
      number of stellar-mass BHs (third and fourth columns) for the
      same simulations shown in the respective top panels.}
    \label{fig:lr1_imbh_nbh}
\end{figure*}

\subsection{Time evolution of the $1\%$ Lagrangian radius}
\label{subsec:survey_sims}
Figures \ref{fig:lr1_cc_nocc} and \ref{fig:lr1_imbh_nbh} show the time
evolution of the 1\% Lagrangian radius ($r_{1\%}$, i.e., the radius
including 1\% of the total cluster mass) for a few selected runs
representative of different dynamical evolutionary histories found in
our survey. The upper panels of Fig.  \ref{fig:lr1_cc_nocc} show two
cases where the initial expansion driven by heavy mass loss from
young, massive stars is followed by a phase where two-body relaxation
becomes dominant, making $r_{1\%}$ to contract progressively and lead
to the CC and post-CC phases. The upper left panel simulation depicts
a system undergoing CC \footnote{ All over the paper, ``CC" refers to the core collapse of the visible component which follows the early collapse of the BH sub-system in the cases where BHs are retained within the cluster.} before 13 Gyr.  Conversely, the upper right panel illustrates the case of a system that experiences CC slightly after 13 Gyr. As discussed in Section \ref{sec:intro}, the simulations with k=1 where BHs are initially retained the clusters are characterized by an early inner contraction due to the early segregation and core collapse of the BH subsystem. The two lower panels of the same figure show the cases
of simulations in which $r_{1\%}$ experiences an extended contraction
phase and, after 13 Gyr of evolution, are still far from CC (which
likely occurs significantly later than 15 Gyr).

Fig. \ref{fig:lr1_imbh_nbh} shows representative cases of simulated
clusters that, at an age of 13 Gyr, host an intermediate-mass BH
(IMBH) or a large population of stellar-mass BHs. The leftmost upper
panel shows $r_{1\%}$ undergoing a contraction phase in the first 6
Gyr, followed by a rapid and significant expansion.  As can be seen in
the panel below, this behavior is due to the presence of an IMBH,
which rapidly starts growing in mass around 6 Gyr, reaching a value of
M$\sim 229 M_\odot$ at 13 Gyr (see \citealp{giersz+15} for a detailed
study of possible evolutionary paths leading to the formation of
IMBHs).  The IMBH becomes a dominant energy source in the core,
sustaining the cluster against contraction and triggering the
expansion of $r_{1\%}$.  The panels in the second column of
Fig. \ref{fig:lr1_imbh_nbh} show a different case of simulations with
an IMBH. Here, the mass growth of the IMBH is more gradual, starting
already from the very beginning of the run (bottom
panel). Correspondingly, $r_{1\%}$ shows an almost constant, slightly
expanding, evolution in time (upper panel).

The remaining panels of the same figure present simulations with
reduced BH kick velocity ($k=1$) that both retain a large number of
stellar-mass BHs for many Gyrs, but show distinct behaviors in
$r_{1\%}$.  The top panel in the third column shows a run where
$r_{1\%}$ is expanding for the first 7 Gyr and then undergoes
contraction due to two-body relaxation (similar to the simulation
named ``DRr'' in B23). As shown in the corresponding bottom panel, the
system loses stellar-mass BHs at considerably high rate, with no BHs
left at an age of 10 Gyr.  The rightmost column of
Fig.~\ref{fig:lr1_imbh_nbh} shows a run with $k=1$ but where $r_{1\%}$
experiences a continuous expansion for the entire cluster lifetime.
The time evolution of the number of BHs in this system (rightmost
bottom panel) indicates that the BH ejection proceeds at a much lower
rate than in the previous case (third column in the Figure \ref{fig:lr1_imbh_nbh}), and $\sim 400$ such objects are still present at an age of 13
Gyr.  The large majority of simulations started with $k=1$ shows
behaviors of $r_{1\%}$ like those plotted in the rightmost panels of
this figure.  In only two instances (simulations
\simu{500}{1}{2}{0.025}{5} and \simu{500}{1}{2}{0.025}{7}; the latter
is the run named ``DRe" in B23), the BH ejection rate is rapid enough
for the systems to undergo a substantial dynamical evolution and reach
CC and post-CC phases.

Altogether, the time dependence of $r_{1\%}$ within the survey
simulations reveals a large variety of dynamical evolutionary
histories. Based on the behavior of $r_{1\%}$, we classify the
simulated cluster extracted at $t=13$ Gyr into three main categories:
``near to CC (nCC)'' if the cluster is reaching CC in the next 1-2 Gyr
(as in the top-right panel of Fig.  \ref{fig:lr1_cc_nocc}), ``CC'' if
they are in CC or post-CC phases (as in the top-left panel of Fig.
\ref{fig:lr1_cc_nocc}), and ``pre-CC'' in all the other
cases. The snapshots showing a rapid growth of an IMBH in their center at the time of CC (as the one shown in the upper-left
panel of Fig. \ref{fig:lr1_imbh_nbh}) are classified as pre-CC clusters because the substantial re-expansion they suffer brings them to have,
at an age of 13 Gyr, the typical structure of poorly evolved systems, with large values of the 1\% Lagrangian radius.

\subsection{Comparison with Galactic GCs}
\label{subsec:comparison_GGCs}
To verify whether the synthetic clusters are representative of the
population of GCs observed in our Galaxy, here we compare their
structural parameters with those listed in the \citet[][2010
  version]{harris96} catalog.

\begin{figure*}[ht]
\centering
    \includegraphics[width=0.33\textwidth]{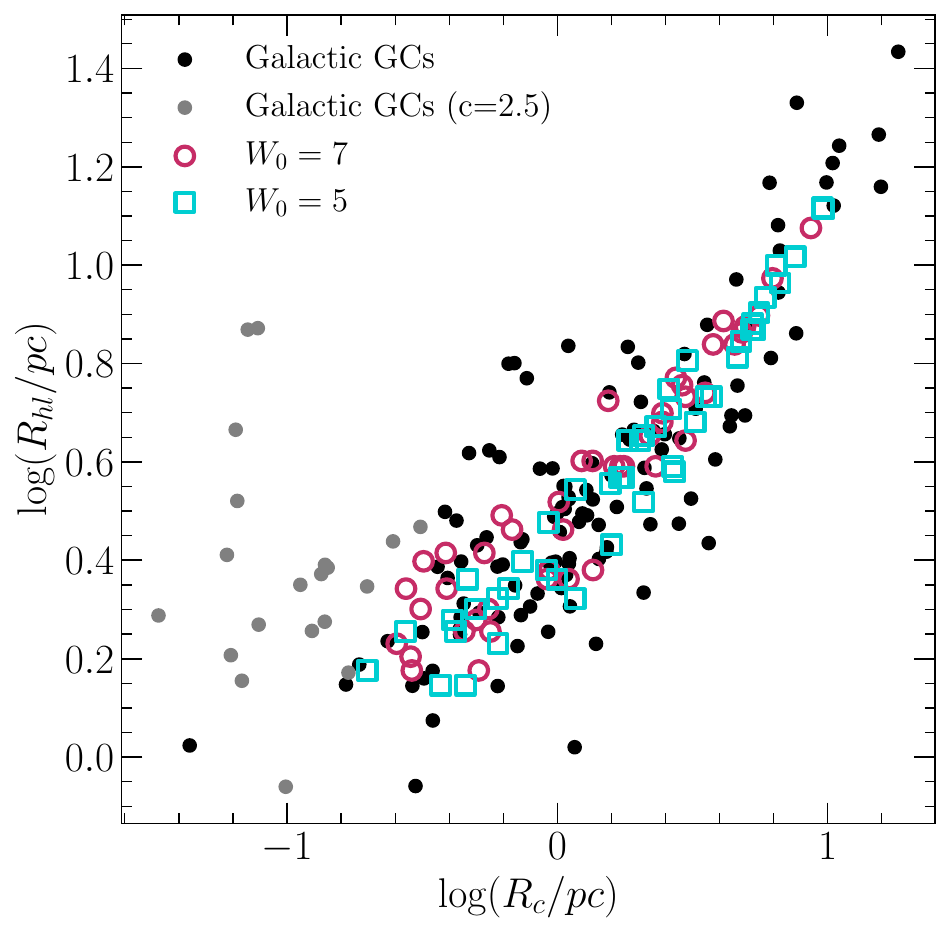}
    \includegraphics[width=0.34\textwidth]{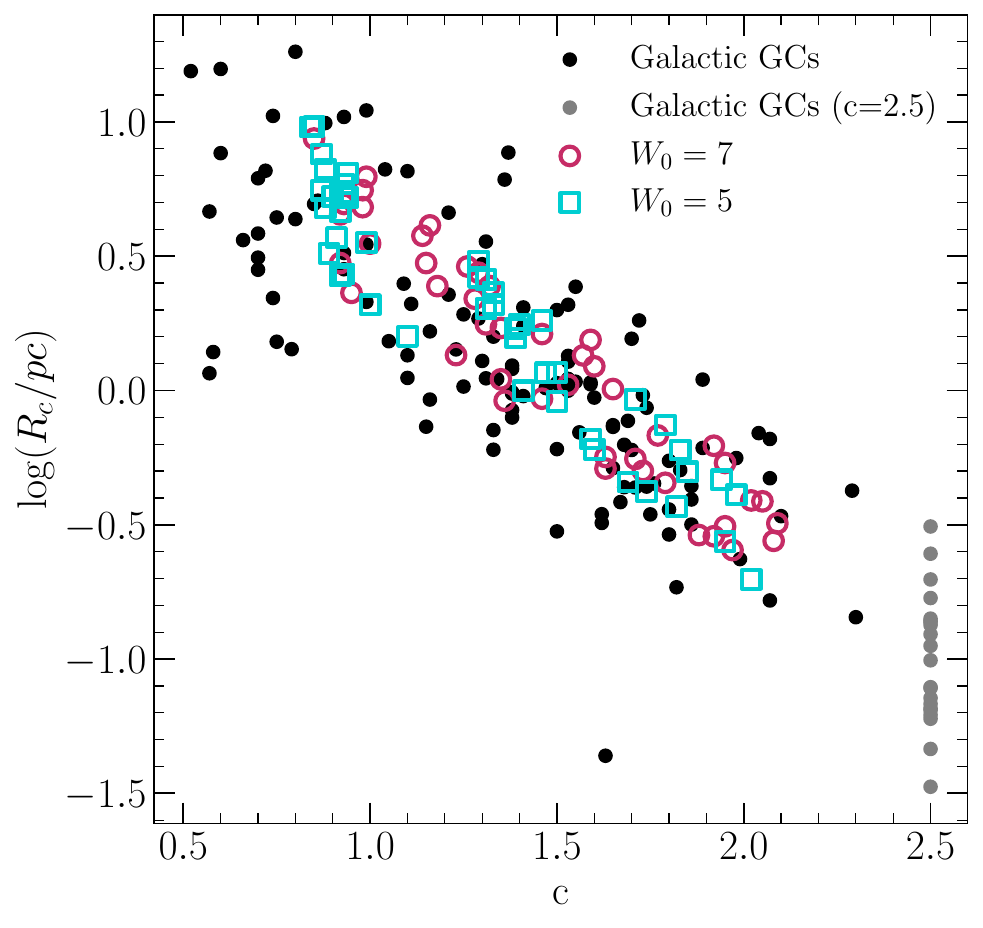}
    \includegraphics[width=0.33\textwidth]{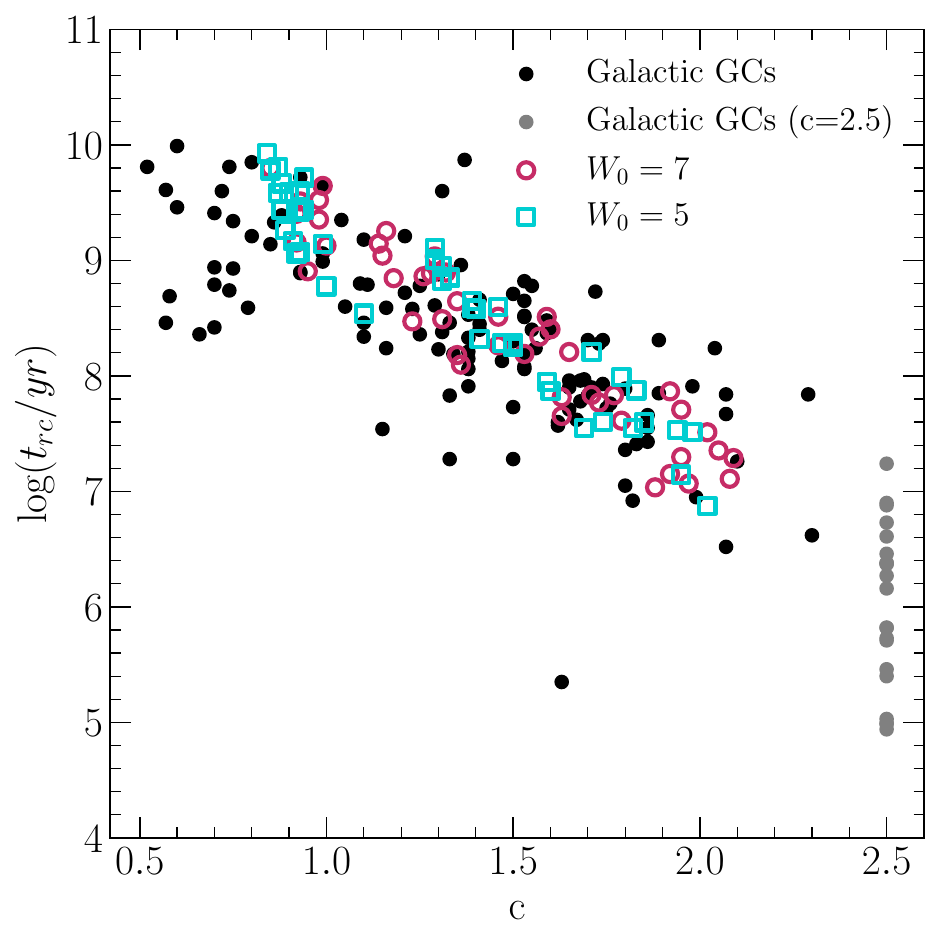}
\caption{Half-light radius ($R_{hl}$) plotted against core radius
  ($R_c$, left panel), core radius plotted against the King
  concentration parameter ($c$, right panel) and core
  relaxation time ($t_{rc}$) plotted against concentration for the 92
  simulation snapshots extracted at 13 Gyr (cyan squares for $W_0=5$,
  and red circles for $W_0= 7$), compared to those measured in the
  Galactic GC population (from \citealt{harris96}),  which are shown as filled circles (gray color for the CC and post-CC systems with c=2.5, black color for the others).
}
\label{fig:c_rc_rhl_trc_harris}
\end{figure*}

To determine the structural parameters of the 92 simulated clusters
extracted at 13 Gyr from our Monte Carlo survey, we follow the same
procedure described in B22 and adopted in observational works \citep[e.g.,][see also
  B22]{miocchi+13, Lanzoni+19}. This consists in determining the
projected density profile from resolved counts of stars brighter than 1 magnitude below the main sequence turnoff point (i.e., with $V< V_{TO}+1$), in a series of
concentric rings around the cluster center, and then search for the
best-fit King model through a $\chi^2$ minimization approach (see also Section 3.1 in
B22 and Section 3 in B23).  We find that the King model family fits quite well the
density profiles of most ($\sim 80\%$) the snapshots, with the
exclusion of the ones that are close to and beyond CC, when a density
cusp develops in the center. As discussed in B22 and B23, for the sake
of homogeneity and reproducibility, and to avoid arbitrariness, the
cusps are not excluded from the King fits.  These density cusps are
shallow compared to those presented in B22, consistent with the fact
the present runs started with 10\% binary fraction and in some cases
retain a significant fraction of stellar-mass BHs (see also B23). From the best-fit King model to each simulated cluster,
  we obtained structural parameters as the core radius $R_c$ (i.e.,
  the radius at which the central projected density is halved), the
  half-light radius $R_{hl}$ (i.e., the radius including half the
  total projected light), and the concentration parameter $c$ (which
  is defined as the logarithm of the ratio between the tidal and the
  King radii). We also determined the core relaxation time ($t_{rc}$)
of each system following the same procedure adopted in observational
studies, i.e., by using equation (10) in \citet{djorgovski93} and
adopting an average stellar mass $<m_*>= 0.3 M_\odot$ and a $V-$band
mass-to-light ratio $M/L_V= 3$.

Figure \ref{fig:c_rc_rhl_trc_harris} shows the resulting trends among $R_{hl}$, $R_c$, $c$, and $t_{rc}$, for the simulated clusters (empty symbols), compared to the values listed in the Harris catalog for the observed Galactic GCs (filled circles). Overall the simulations occupy the same regions of the parameter space covered by the observations. The only exception concerns the regions occupied by the Galactic GCs classified as CC and post-CC in the Harris compilation (gray filled circles in the figure), where no simulated systems are found. This discrepancy is due to the different approach adopted for the determination of the structural parameters of these clusters, which show a central cusp in their density profile: while we search for the best-fitting King model to the entire density profile (including the central cusp), in the Harris catalog the concentration parameter is arbitrarily fixed to a constant value (c=2.5) and $R_c$ is determined as the radius where the surface brightness is equal to half the central value \citep[see][]{trager+93}. Hence, for the purposes of our analysis, the simulated clusters at 13 Gyr can be considered well representative of the Galactic GC population.

\subsection{Normalised cumulative radial distributions and nCRD parameters}
\label{sec:construction_ncrd_paper3}

\begin{figure*}[ht]
    \centering
    \includegraphics[width=0.7\textwidth]{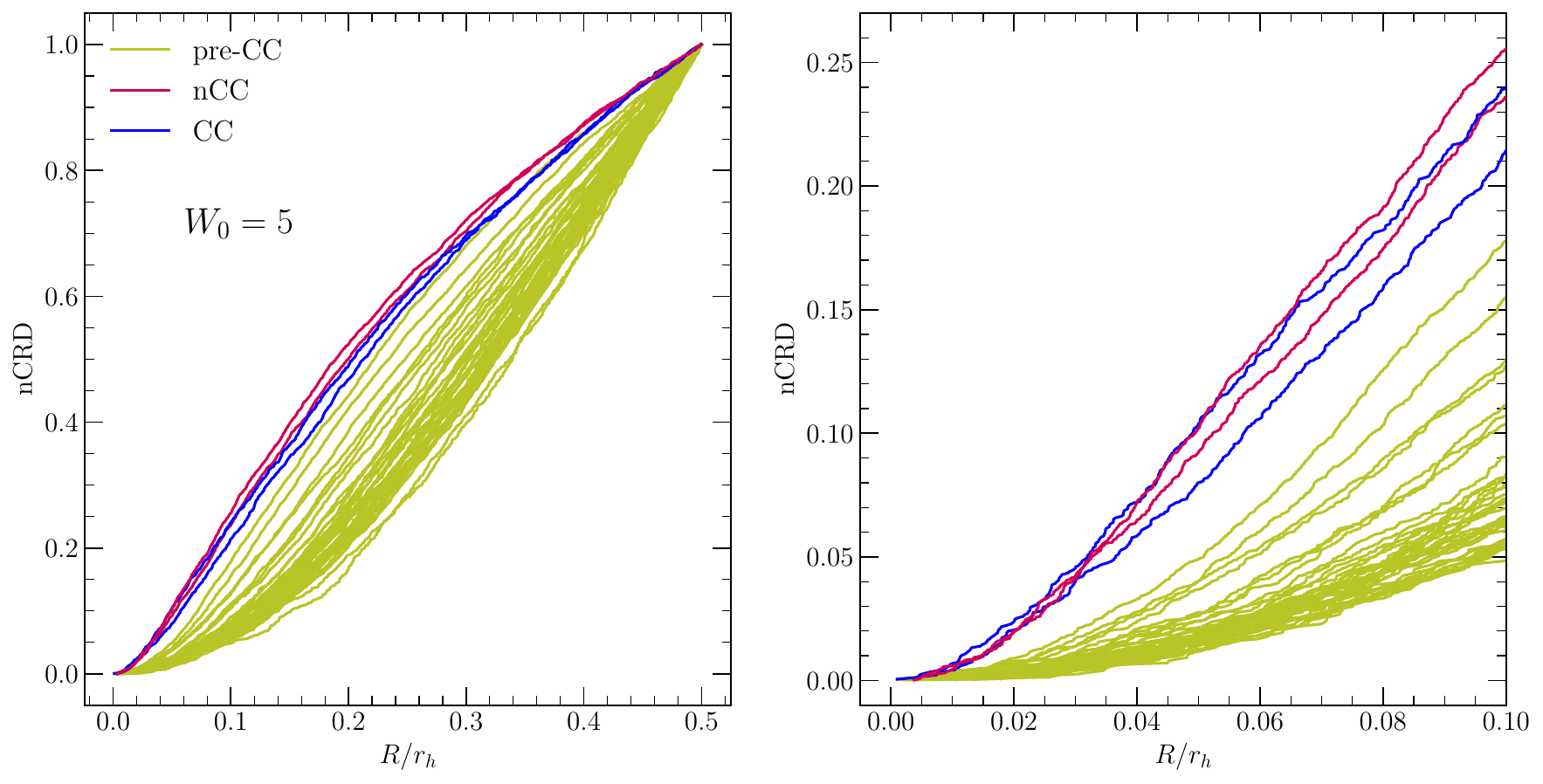}
    \caption{Left panel: Normalized cumulative radial distributions
      (nCRDs) of the stars brighter than $V_{\rm TO}+0.5$ and located
      within 0.5$\times r_h$ from the center for the snapshots
      extracted at 13 Gyr from the $W_0=5$ runs. Right panel: zoom
      into the inner cluster region. The color code is as follows:
      green for dynamically young (pre-CC) snapshots, red for
      snapshots close to CC (nCC), and blue for dynamically-old
      snapshots.}
    \label{fig:crd_w05}
\end{figure*}

Once the value of $r_h$ is obtained from the King model fit to the projected density profile, to determine the new diagnostics of dynamical evolution defined in B22
and B23, the first step is to build the nCRDs of all the stars with $V< V_{\rm TO} + 0.5$, and located
within a projected distance ($R$) equal to $0.5 \times r_h$ from the
center, $r_h$ being the three-dimensional half-mass radius of the single-mass King model fitting the projected density profile (built by using stars with $V<V_{TO}+1)$.

\begin{figure*}[ht]
    \centering
    \includegraphics[width=0.7\textwidth]{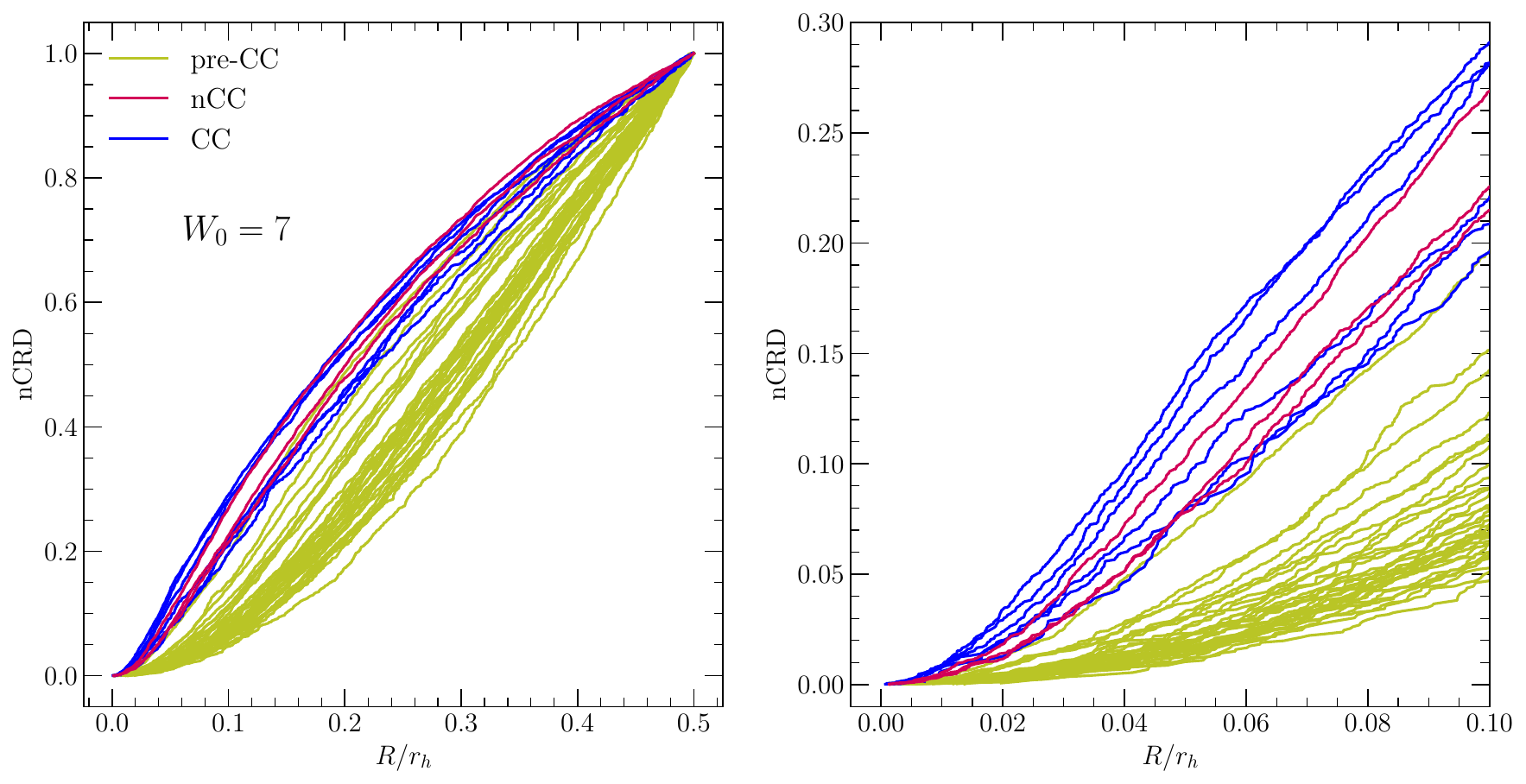}
    \caption{As in Fig. \ref{fig:crd_w05}, but for the snapshots
      extracted from the $W_0=7$ runs.}
    \label{fig:crd_w07}
\end{figure*}

Figures \ref{fig:crd_w05} and \ref{fig:crd_w07} show the nCRDs of all
the analyzed snapshots started with $W_0=5$ and $W_0=7$, respectively,
with the right panels zooming into the innermost cluster region ($R
<0.1 r_h$). The color code is based on the dynamical evolutionary
stage identified for each extracted snapshot from the respective
$r_{1\%}$ evolution (see caption).  The dynamical evolution
classification based on $r_{1\%}$ is well mirrored by different shapes
of the nCRDs, with dynamically younger systems showing systematically
shallower curves, i.e., by construction, smaller percentages of stars
included within the inner cluster regions. This is well in agreement
with the expected effect of internal dynamical evolution, which tends
to progressively increase the central density of star clusters,
therefore steepening their nCRDs. The only exception (see
  the green line close to the red and blue ones in the right panel of
  Fig. \ref{fig:crd_w07}) seems to be simulation
\simu{1}{0}{6}{0.025}{7} that is classified as pre-CC because, based
on its $r_{1\%}$ evolution, shows no evidence of being close to CC,
but has a nCRD ``shifted within the group of highly evolved systems''.
A closer inspection of the figure, however, shows that this nCRD is
the least steep within this group. This likely indicates that the
system is more dynamically evolved than all the other ones classified
as pre-CC, but still in an earlier phase than what we labelled
``nCC''.

The three parameters presented in B22 and B23 have been specifically
defined to quantify the observed morphological changes of the nCRD,
which, in turn, are tracers of the dynamical age of the systems. Here
we briefly recall their definition:
\begin{itemize}
\item $A_5$ is the area subtended by each nCRD between the center and
  a projected distance equal to 5\% $r_h$ ($R/r_h=0.05$);
\item $P_5$, is the value of the nCRD (i.e., the fraction of stars) at
  the same distance from the center;
\item $S_{2.5}$ is the slope of the straight line tangent to the nCRD
  at $R/r_h=0.025$. More specifically, it is the slope of the tangent
  to the third-order polynomial function that best-fits the nCRD.
\end{itemize}
Adopting the above definitions, we determined the values of $A_5$,
$P_5$, and $S_{2.5}$ for all the snapshots extracted at 13 Gyr from
our Monte Carlo simulation survey. Out of 92 snapshots, two have less
than ten stars inside their 5\% $r_h$, thus preventing any meaningful
measure of the nCRD parameters. Hence, they have been discarded and
only 90 snapshots have been used in the following analysis.

\section{Results}
\label{sec:results}
\subsection{Reference boundaries and models from previous works}
\label{sec:king_sequence_cc_boundary}
For the proper interpretation of the present results, we take
advantage of our previous works.  As discussed in B23, star clusters
in different dynamical phases tend to occupy different regions of the
$A_5-P_5$, $A_5-S_{2.5}$, and $P_5-S_{2.5}$ diagrams, irrespective of
the primordial binary fractions and dark remnant content.  We
therefore exploit those results to identify regions in these diagrams
where dynamically-young and dynamically-old systems are expected to
lie.

\begin{figure*}[htb]
    \centering
    \includegraphics[width=0.7\textwidth]{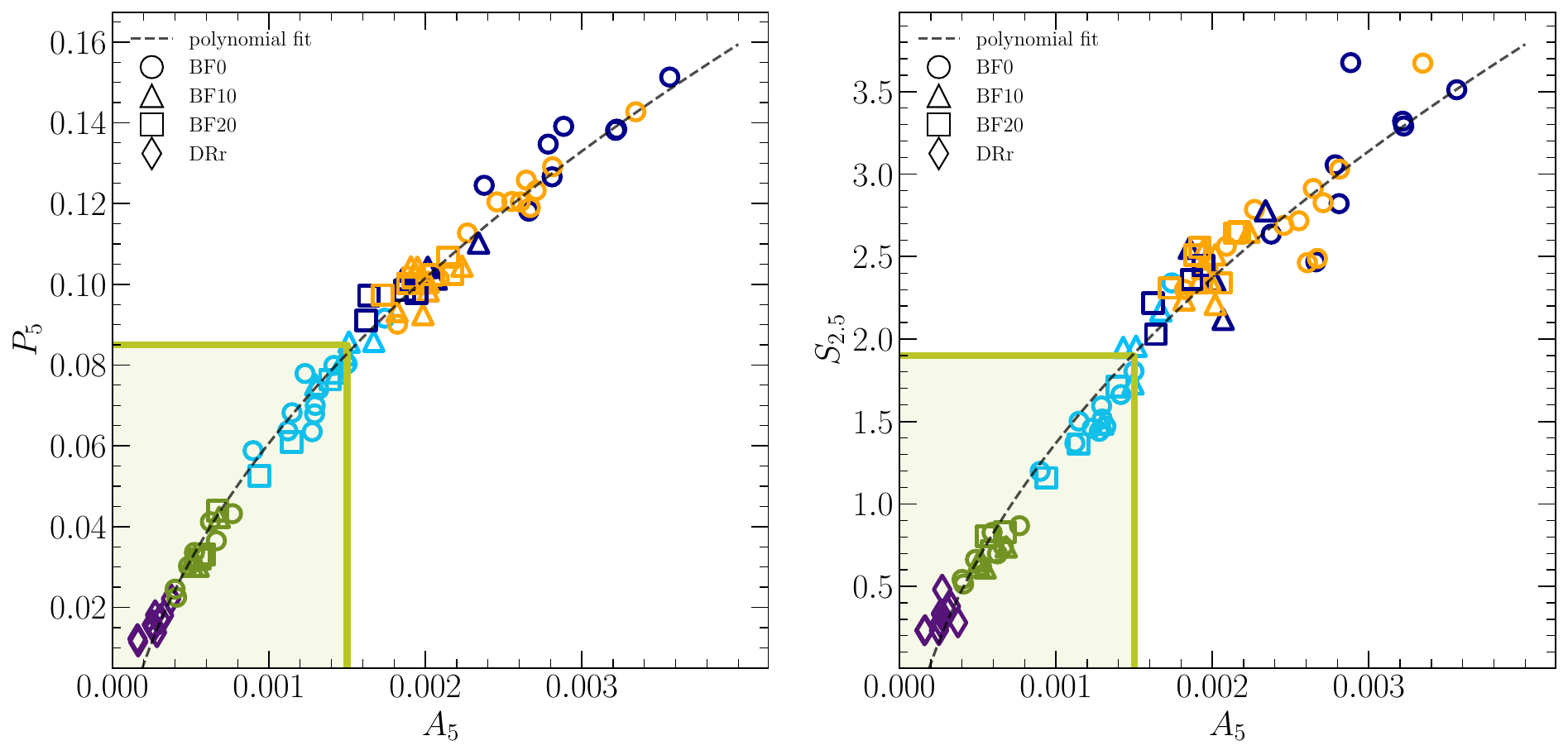}
    \caption{$P_5$ and $S_{2.5}$ parameters plotted against $A_5$
      (left and right panels, respectively) for the reference
      simulations discussed in B22 and B23, started with 0\%, 10\% and
      20\% binary fractions (BF0, BF10, and BF20, respectively) and
      retaining a large population of dark remnants for the entire
      time evolution (DRr).  The color code of the symbols is as
      follows: green, cyan, blue, and orange for early, intermediate,
      CC, and post-CC phases, respectively; indigo diamonds refer to
      the DRr simulation, which always remains in very early phases of
      dynamical evolution.  The boundary region including only
      dynamically-young (pre-CC) systems is shaded in green.  The
      polynomial fits to the points are plotted as black dashed
      lines.}
    \label{fig:a5_p5_boundary}
\end{figure*}

In Figure \ref{fig:a5_p5_boundary} we plot the values of the three
parameters measured in B22 and B23 for simulations with 0\%, 10\%,
20\% primordial binary fractions (BF0, BF10, and BF20, respectively),
and for a case where a large amount of BHs was retained within the
cluster for its entire evolution (DRr). Different colors correspond to
different dynamical stages: green, cyan, blue and orange for early,
intermediate, CC, and post-CC stages, respectively. Indigo is used for
the DRr simulation that shows no evolution from the dynamical point of
view, due to heating effects of the retained BHs.  It is evident that
the snapshots that are in early and intermediate dynamical stages, as
well as the snapshots of the DRr simulation, all occupy the lower-left
part of the plots. Conversely, the snapshots in CC and post-CC phases
occupy the upper-right part of the diagrams. There is also a small
overlapping region where a few intermediate and highly evolved
snapshots are found to lie together. Based on this evidence, we drew a
boundary region (green shaded) safely including only dynamically-young
snapshots from all the simulations analyzed in previous works.
Specifically, the boundaries for pre-CC systems are set at $A_5\leq$
0.0015, $P_5\leq$ 0.085 and $S_{2.5}\leq$ 1.9.  Although these
boundaries are somehow arbitrary, they serve as reference and guide
for the interpretation of the results obtained from the present
simulation survey.  For the same purposes we also fitted a polynomial
function to the sequences of $A_5$ versus $P_5$, and $A_5$ versus
$S_{2.5}$ (black dashed lines in Fig. \ref{fig:a5_p5_boundary}).

\subsection{Effects of different initial conditions}
\label{sec: ncrd_param_nkrgrp}
The dynamical evolution of star clusters is known to be affected by
the {\it combination} of several different processes and cluster
properties. Here, we take advantage of the simulation survey to search
for possible dependences of the results on each initial condition
parameter considered individually, namely, the total number of stars
$N$, the BH kick velocity prescription ($k$), the galactocentric
distance ($R_g$), the tidal filling parameter (TF), and the King
dimensionless potential ($W_0$; see Table \ref{tab:list of
  sims_paper3}).

\begin{figure*}[ht]
\centering
    \includegraphics[width=0.9\textwidth]{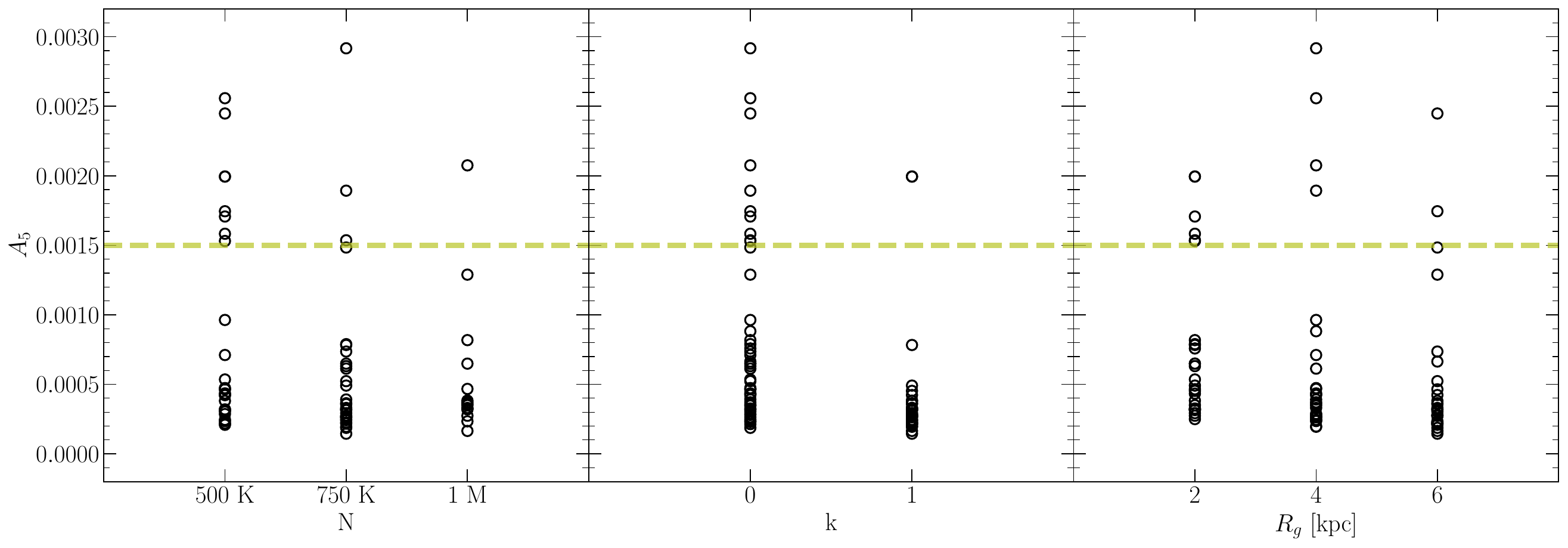}
    \includegraphics[width=0.6\textwidth]{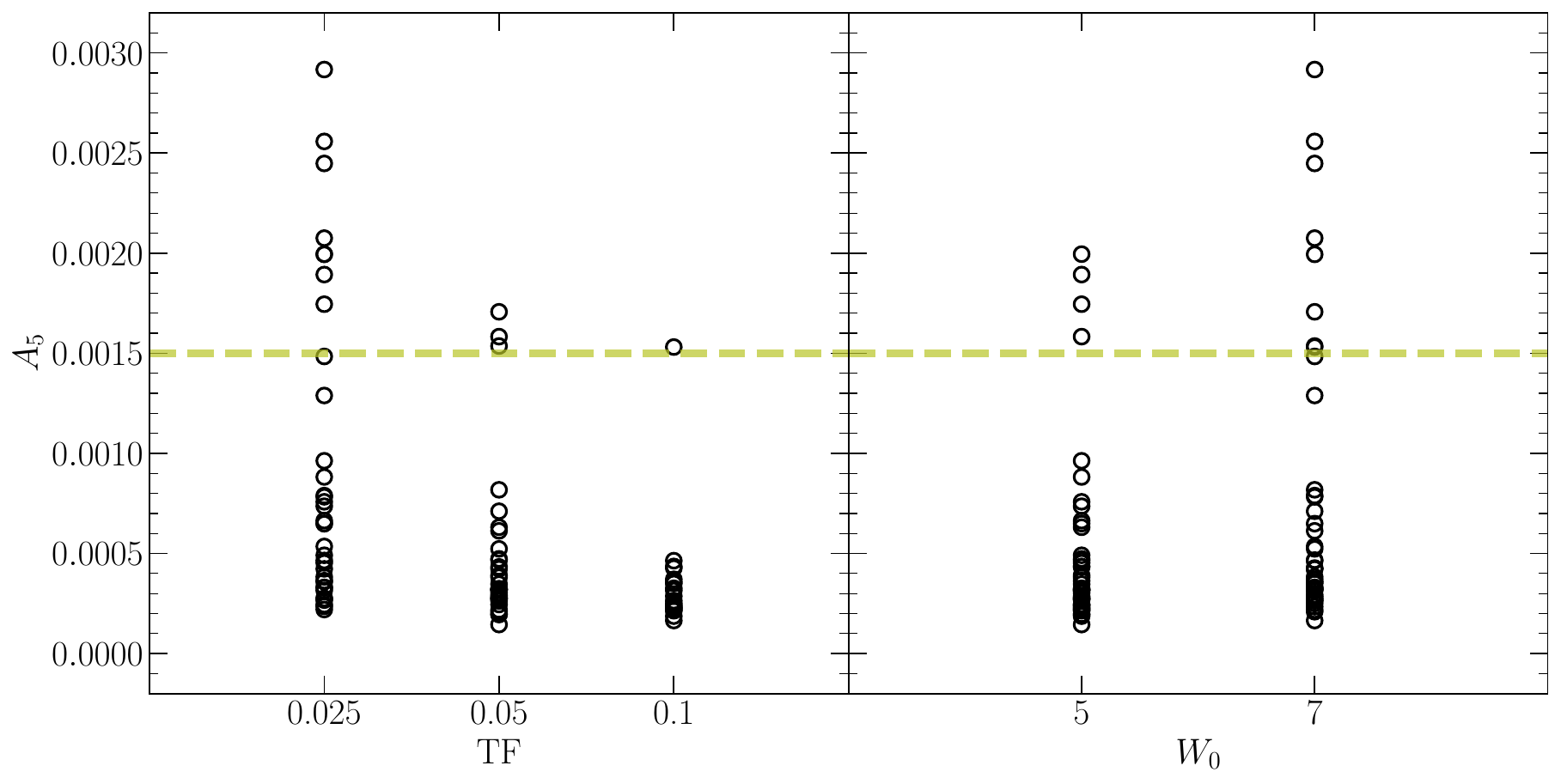}
    \caption{$A_5$ parameter measured in the analyzed snapshots
      plotted as a function of the adopted initial conditions (see
      Table \ref{tab:list of sims_paper3}): from top-left, to
      bottom-right the total number of stars $N$, the BH kick velocity
      prescription ($k$), the galactocentric distance ($R_g$), the
      tidal filling parameter (TF), and the initial King dimensionless
      potential ($W_0$). The dashed lines mark the value of $A_5$
      adopted as boundary to separate clusters in pre-CC and more
      advanced dynamical phases ($A_5 = 0.0015$).}
\label{fig:ics}
\end{figure*}

Figure \ref{fig:ics} summarizes the results obtained for the $A_5$
parameter. These are perfectly analogous to those found for $P_5$ and
$S_{2.5}$, which are therefore not shown here. The dashed lines
correspond to the value $A_5 = 0.0015$ that, based on our previous
works, marks the separation between systems that are in early and in advanced dynamical stages.  As a general consideration, the majority of the
snapshots are found in pre-CC dynamical phases (i.e., show $A_5 \le
0.0015$), while only a few (13 out of 90) are in the dynamically-old
region (above the dashed line in the figure). This is consistent with
the shapes of the nCRDs discussed in Section
\ref{sec:construction_ncrd_paper3}, most of which are in the ``shallow
group'' (green lines). The top-left panel of Fig. \ref{fig:ics} shows
a mild indication that the most massive systems ($N=$ 1M) do not reach
the highest values of $A_5$ (corresponding to the highest levels of
dynamical evolution), compared to simulations started with lower
values of $N$.  This indicates that, as expected, $N$ plays an
important role in the clusters' dynamical history; for a given initial
spatial distribution, the half-mass relaxation time increases with the
cluster mass, and thus the dynamical evolution proceeds on a longer
timescale leading to the trend shown in this panel.  A hint of more
advanced dynamical stages for larger fractions of systems with small
initial Galactocentric distances ($R_g=$ 2 and 4 kpc, compared to
$R_g=6$ kpc) is observable in the top-right panel, while more
significant trends are found between $A_5$ and the remaining
parameters. Faster dynamical evolutions (i.e., larger values of $A_5$
and larger percentages of snapshots above the dashed lines) are
observed for systems that starts with smaller BH retention ($k=0$,
compared to $k=1$), higher initial compactness (i.e., smaller degree
of tidal underfilling: TF=0.025, compared to TF=0.05 and 0.1), and
higher concentration ($W_0=7$, compared to $W_0=5$).

In particular, while the snapshots with $k=0$ cover the entire range
of $A_5$ values, those with reduced BH kick velocity ($k=1$) are
essentially all clumped in the low-end of the distribution,
corresponding to young dynamical evolutionary states.  This is due to
the effects of retained BH populations, which act as
  energy source that powers core expansion
  \citep[e.g.,][]{heggiehut2003}, delaying the contraction of the
inner region and extending the time needed to reach CC.  The only two
$k=1$ snapshots that fall in dynamically-old region are those already
discussed in Section \ref{sec:analysis} (\simu{500}{1}{2}{0.025}{5}
and \simu{500}{1}{2}{0.025}{7}). They show essentially the same value
of $A_5$ (0.001995 and 0.001994, respectively) and their relatively
fast dynamical evolution is due to the fact that all the BHs are
ejected within the first few Gyr, thus leaving to the cluster inner
regions enough time to substantially contract and reach CC within 13
Gyr.  It is interesting to point out that values of $A_5$
corresponding to dynamically-young clusters (in a pre-CC phase) are
not exclusively present in models retaining BHs (i.e., models with
$k=1$) but also in a large number of models with no BH retention (with
$k=0$).  A strong impact on $A_5$ is also observed from the TF
parameter: only simulations that started with TF= 0.025 (the most
compact in the survey) reach the dynamically-old region in the plot,
while for higher values of TF, almost all the snapshots have $A_5 \le
0.0015$ and several clusters with TF=0.1 dissolve within 15 Gyr of
evolution.

\subsection{nCRD parameters as diagnostics of dynamical evolution}
\label{sec:ncrd_diagnostics_paper3}

\begin{figure*}[ht]
\centering
    \includegraphics[width=\textwidth]{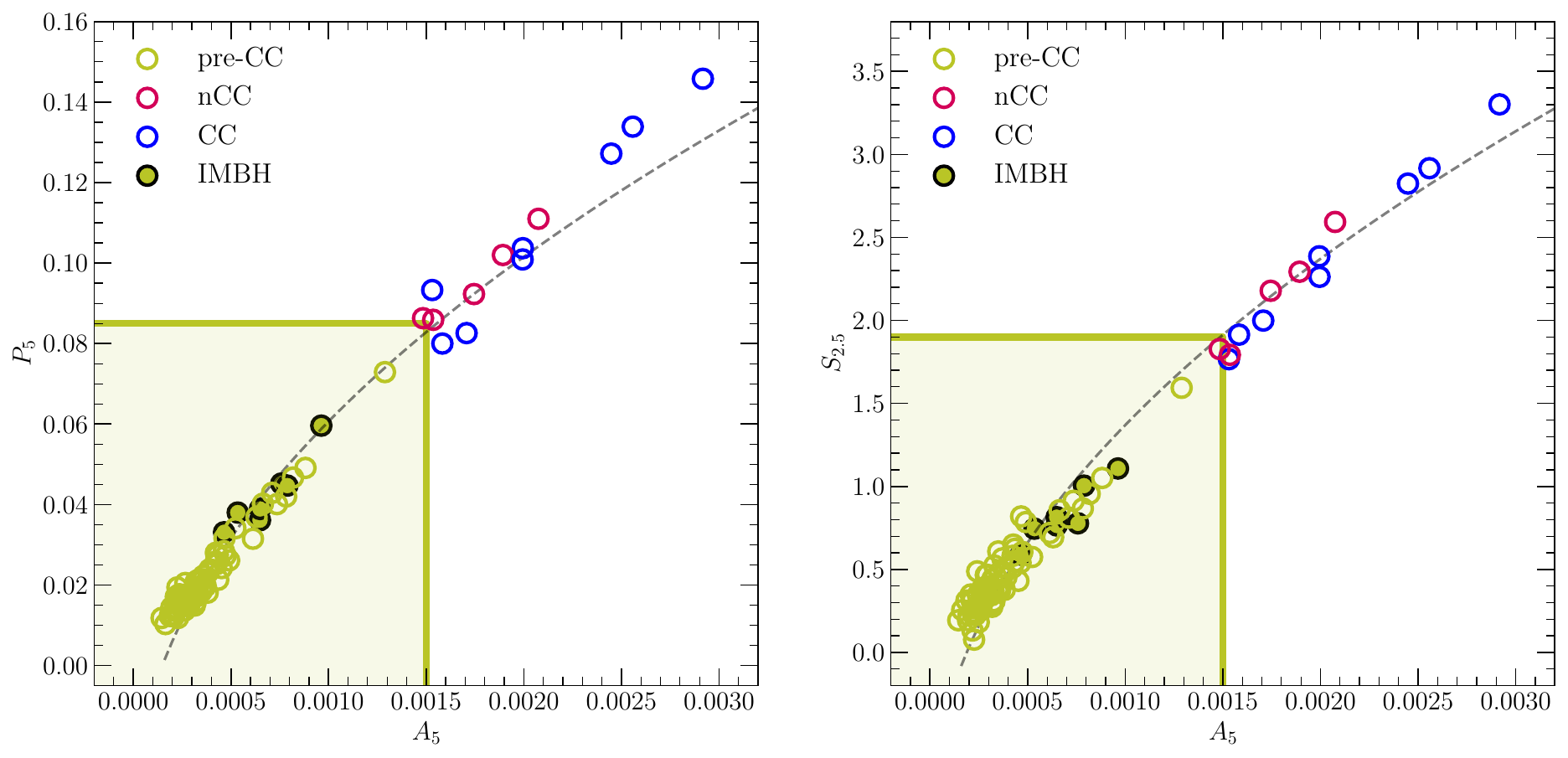}
     \caption{$P_5$ and $S_{2.5}$ parameters plotted against $A_5$
       (left and right panels, respectively), as measured in the 13
       Gyr snapshots of the simulation survey discussed in this work.
       The color code is the same as in Figs. \ref{fig:crd_w05} and
       \ref{fig:crd_w07}, indicating different dynamical phases: green
       for dynamically-young snapshots (pre-CC), red and blue for
       systems in the verge to collapse (nCC) and that already reached
       CC, respectively.  The snapshots with IMBH are encircled in
       black.  The green shaded regions and the black dashed lines
       are, respectively, the reference boundaries for pre-CC systems
       and the polynomial fits to the simulations discussed in B22 and
       B23, as defined in Section \ref{sec:king_sequence_cc_boundary}
       (the same as in Fig.  \ref{fig:a5_p5_boundary}).}
\label{fig:a5_p5_dynamical_stages}
\end{figure*}

The adopted reference boundaries for pre-CC systems (green shaded
regions in Figure \ref{fig:a5_p5_boundary}) have been chosen based on
snapshots extracted at different chronological times during the
cluster evolution, from the simulations discussed in B22 and B23.
Conversely, in the present work the $A_5$ , $P_5$ and $S_{2.5}$
parameters have been all measured in snapshots extracted at the same
age (13 Gyr) from cluster simulations run from different initial
conditions. Hence, it is not obvious \emph{a priori} that the current
results closely follow the previous ones.

To verify this possibility, Figure \ref{fig:a5_p5_dynamical_stages}
shows the values of $P_5$ and $S_{2.5}$ as a function of $A_5$ (left
and right panels, respectively) measured from the simulation survey
discussed in this work, compared to the boundary region for pre-CC
clusters defined from B22 and B23 (green shaded area) and the
polynomial functions fitted to our previous results (dashed lines; see
Sect. \ref{sec:king_sequence_cc_boundary}).  The present results
(empty circles) are colored as in Figs.  \ref{fig:crd_w05} and
\ref{fig:crd_w07} based on the evolution of the 1\% Lagrangian radius,
with green color for the pre-CC snapshots, red color for the clusters
close to CC, and blue for those that already reached CC at 13 Gyr.  As
apparent, they are all aligned along narrow sequences that closely
follow the polynomial fits to the results of B22 and B23.  In
addition, the green shaded region is populated exclusively by
snapshots classified as pre-CC, while all the dynamically-old systems
lie outside this boundary.  Also the simulations classified as nCC
show values outside the reference boundary for pre-CC snapshots,
indicating their higher level of dynamical evolution, which is also
testified by the shape of their nCRDs, despite their 1\% Lagrangian
radius did not formally reach CC at 13 Gyr yet.  Although a clear
distinction among the different stages of high dynamical evolution
(nCC, CC, post-CC) is not possible from the nCRD parameters, this
figure clearly demonstrates the suitability of the new diagnostics and
the adopted boundaries to substantially distinguish between
dynamically-young and dynamically-old snapshots.

The green circle close to the top-right corner of the boundary region
is simulation \simu{1}{0}{6}{0.025}{7}, the one with the nCRD
``shifted within the group of highly evolved systems'' (right panel of
Fig. \ref{fig:crd_w07}). Its position in this diagram suggests that
this system is more evolved than all the other pre-CC clusters, fully
confirming the conclusions drawn in Section
\ref{sec:construction_ncrd_paper3}. In turn, this demonstrates that the
nCRD parameters have higher sensitivity to dynamical evolution than
$r_{1\%}$, which is unable to quantify any difference between this
case and the other pre-CC ones.


\begin{figure*}[ht]
    \centering
    \includegraphics[width=0.5\textwidth]{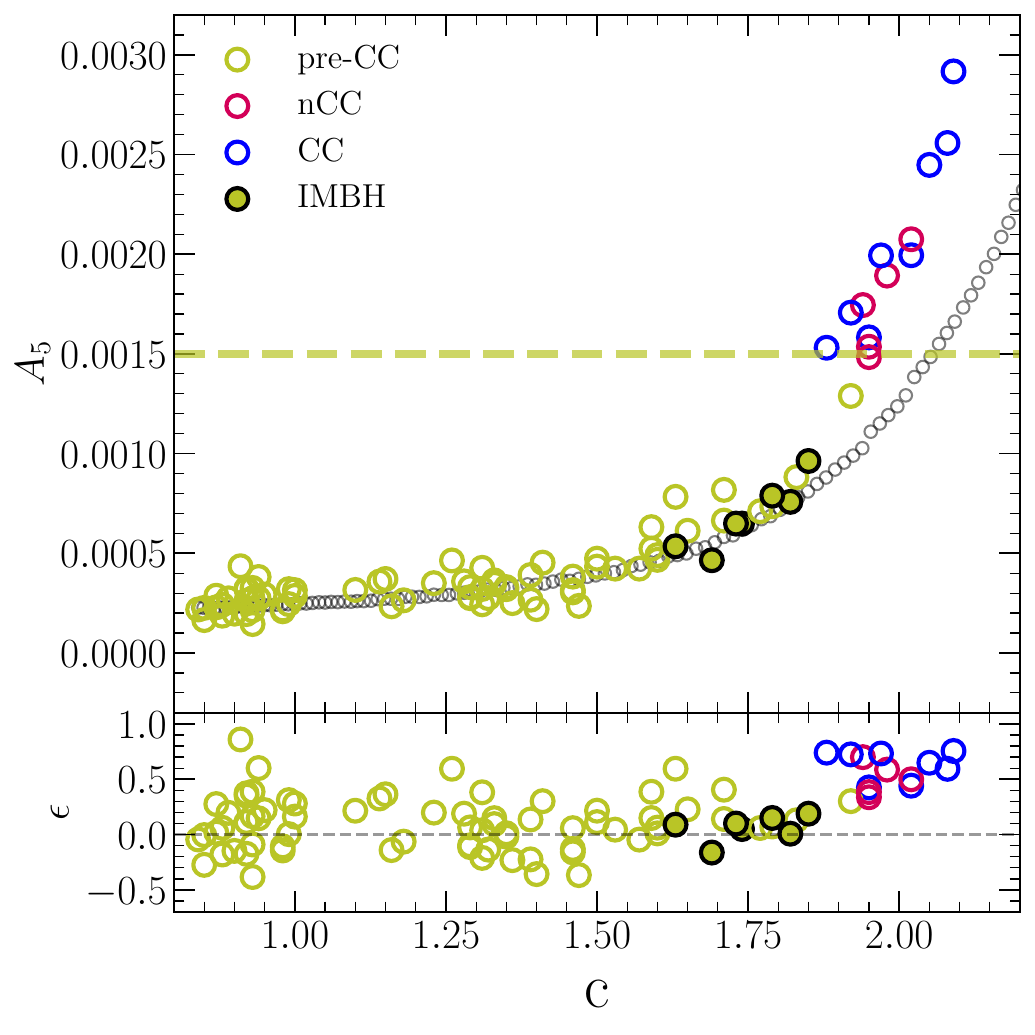}
    \caption{$A_5$ measured in the survey simulation snapshots (large
      circles) plotted against their concentration parameter as
      obtained from the best-fit King model to their density
      profile. The color code is the same as in
      Fig. \ref{fig:a5_p5_dynamical_stages}, and the green dashed line
      marks the separation boundary between pre-CC and more evolved
      clusters. The small grey circles are the values of $A_5$
      obtained by directly integrating the density profile of King
      models with concentration $c$ varying in the range 1 and 2.5, in
      steps of 0.05. The bottom panel shows the residuals between the
      measured nCRD parameter and the value expected for the King
      model with same concentration.}
    \label{fig:a5_king}
\end{figure*}

Figure \ref{fig:a5_king} shows $A_5$ plotted against the King
concentration parameter ($c$) determined from the King model fitting
to the density profile of each analyzed snapshot (colored
circles). For reference, the small grey circles are the values of
$A_5$ obtained by integrating the density profile of a sequence of
King models with concentration parameter $c$ varying between 1 and
2.5, in steps of 0.05.  As apparent, the values obtained for the
dynamically young snapshots (green circles) well follow the sequence
expected for King models up to c$\simeq 1.8$, with some scatter. Then,
the values of $A_5$ measured in dynamically-old systems (red and blue
circles) systematically and significantly exceed those expected for a
King model with the same concentration $c$.  This is in qualitative
agreement with the well-known behavior of the density profile, which
significantly deviates from the King model distribution (due to the
presence of a central power-law cusp) in advanced stages of dynamical
evolution.  Very interestingly, also the pre-CC cluster that is close
to the boundary between dynamically-young and dynamically-old systems
(simulation \simu{1}{0}{6}{0.025}{7}) shows a significant deviation
from the King model sequence at $c>1.8$, further confirming that it is
more dynamically evolved than its pre-CC peers.

The systems hosting a central IMBH tend to have larger King
concentrations (between 1.63 and 1.85) than most of the other pre-CC
clusters, but closely follow the King sequence. This indicates that
their density profile shows no significant deviations with respect to
the King model behavior. Indeed, previous works \citep[e.g.,][but see
  also \citealp{Vesperini+10}]{baumgardt+05, miocchi07} predict that a
central IMBH should induce a cusp in the innermost portion of the
density profile, with a slope that is much shallower than the one
expected in CC systems.


\section{Summary and Discussion}
\label{sec:discussion_chap4}
In this paper we have measured the three nCRD parameters proposed by
B22 and B23 (namely, $A_5$, $P_5$ and $S_{2.5}$) in a sample of 90 GCs
simulated through the MOCCA Monte Carlo code, starting from different
initial conditions (see Table \ref{tab:list of sims_paper3}). The
simulated clusters have been analyzed following the methods commonly
implemented in observational studies.  The crucial difference with
respect to our previous studies, is that in B22 and B23 the parameters
were measured for just a few models at different evolutionary times
corresponding to different chronological ages (and therefore with
snapshots including stars spanning different mass ranges), while here
we consider a much broader range of initial conditions and compare all
the systems at the same chronological age (13 Gyr). Thanks to the
different intial conditions these systems have reached different
dynamical evolutionary stages, as clearly testified by the variety of
shapes observed for the time evolution of their 1\% Lagrangian radii
(see a few examples in Figs. \ref{fig:lr1_cc_nocc} and \ref
{fig:lr1_imbh_nbh}), and the sample therefore is more appropriate to
represent the Galactic GC population, where all systems are coeval and
old, with ages of 10-13 Gyr, but are characterized by distinct
dynamical ages \citep[see, e.g.,][] {ferraro+18}.  Indeed, the
comparisons between the structural parameters (King concentration,
core and half-light radii) and the core relaxation time of the
synthetic and the observed clusters show that these simulations are in
general well representative of the Galactic GC population.

We find that the three nCRD parameters measured in the present sample
are fully consistent with those obtained in B22 and B23, drawing the
same sequences in the $P_5$ and $S_{2.5}$ versus $A_5$ diagrams, and
with the pre-CC systems showing systematically smaller values than the
more dynamically evolved clusters
(Fig. \ref{fig:a5_p5_dynamical_stages}).  The dependence the nCRD
parameters on the initial conditions of the simulations (see
Fig. \ref{fig:ics} for the case of $A_5$) shows that, as expected,
clusters that form more massive and less compact (i.e., those having
longer initial half-mass relaxation times) tend to experience a slower
dynamical evolution.  This is also the case for clusters retaining
larger populations of BHs until the present time.  Some compact
clusters (specifically the models \simu{500}{1}{2}{0.025}{5} and
\simu{500}{1}{2}{0.025}{7}) retain a large fraction of BHs initially,
but these are then dynamically ejected leaving few or no BHs at the
present time; in those cases the early ejection of BHs leaves enough
time for the cluster to progressively contract and approach CC under
the effect of two-body relaxation (see also \citealp{weatherford+18,
  kremer+20}, B23).  The present-day properties thus depend on the
interplay between initial structural properties and their effects on
the clusters' stellar content, and dynamical energy sources such as
BHs and binary stars.  The clusters classified as dynamically-old show
large King concentration ($c>1.8$) and nCRD parameters that are
systematically larger than the values expected from the direct
integration of the King models computed with the values of $c$
providing the best fit to their density profiles (see
Fig. \ref{fig:a5_king}).  This is in agreement with the
well-established presence of a power-law cusp (i.e., a significant
deviation from the King model behavior) in the central portion of the
density profile of highly evolved GCs.

\begin{figure}[h!]
\centering
    \includegraphics[width=0.45\textwidth]{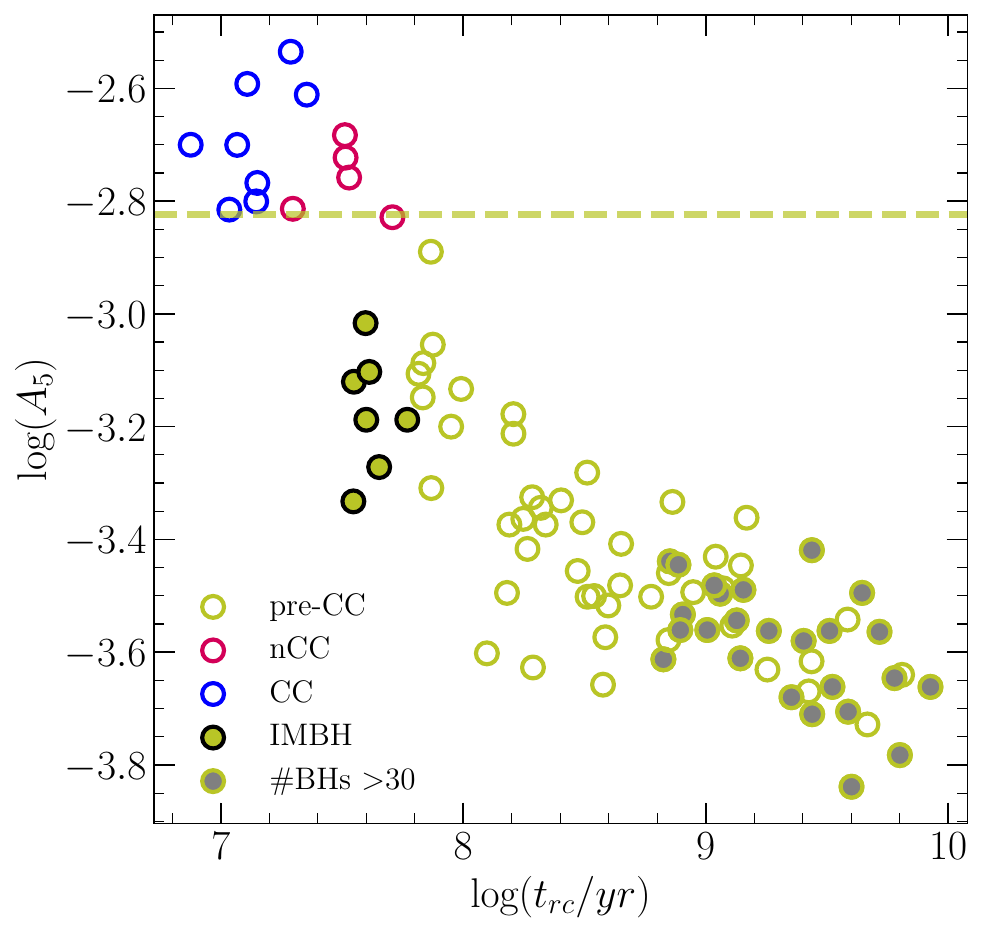}
    \caption{Logarithm of $A_5$ plotted against the logarithm of
      $t_{rc}$.  The symbols are as in Fig. \ref{fig:a5_king} (also
      see the legend), with the addition of the green circles filled
      in grey that correspond to clusters hosting more than 30
      stellar-mass BHs at an age of 13 Gyr.}
\label{fig:trc_a5}
\end{figure}

A further demonstration of the effectiveness of the nCRD parameters as
diagnostics of internal dynamical evolution is provided in Figure
\ref{fig:trc_a5}, which shows a clear anti-correlation between $A_5$
and the present-day value of the central relaxation time
($t_{rc}$). It is interesting to note that nCC clusters are better
distinguishable from CC systems in this plot, compared to
Figs. \ref{fig:a5_p5_dynamical_stages} and \ref{fig:a5_king}. In fact,
for the same value of $A_5$ they show systematically larger values of
$t_{rc}$, consistent with their lower dynamical evolution.  In
addition, $A_5$ is able to distinguish nCC clusters from pre-CC
snapshots (above and below the threshold, respectively) that, showing
essentially the same values of $t_{rc}$, would not be separable in
based on their central relaxation time.  Very interestingly, the
snapshots hosting an IMBH in their center are all clustered in the
upper-left end of the pre-CC sequence in Fig. \ref{fig:trc_a5}, at
$0.00045<A_5<0.001$ and $\log(t_{rc})\sim 7.5$. The IMBHs have masses
ranging between $0.07\%$ and $0.45\%$ of the total cluster mass (i.e., between $\sim 100$ and $1000 M_\odot$), and they
are bound in binary systems with a neutron star, a white dwarf or a
non-degenerate star, but in all cases no other BH is present in the
system. Instead, all the systems currently hosting large populations
of stellar-mass BHs (more than $\sim 30$) are found at much larger
values of $t_{rc}$ and smaller values of $A_5$ (green circles filled
in grey in Fig. \ref{fig:trc_a5}).  This suggests that the heating
effect of a few hundred solar mass IMBH alone tends to be weaker than
that of several stellar-mass BHs, and this result might help breaking
the degeneracy between the presence of a single massive compact object
and a concentration of dark remnants in the center of a GC, which
often remain as possible indiscernible alternatives to explain an
observed rising of the velocity dispersion in the innermost cluster
region (see, e.g., \citealp{vandenbosch+06} for the case of M15 and
\citealp{ vitral+23} for M4).

All these findings prove that the three nCRD parameters $A_5$, $P_5$
and $S_{2.5}$ are powerful diagnostics of the stage of internal
dynamical evolution reached by star clusters.  Compared to the
computation of $t_{rc}$, they offer the advantage of being fully
empirical and not relying on simplifying assumptions and
approximations.  With respect to alternative methods requiring the
determination of the stellar mass function or internal kinematics at
different radial distances from the center
\citep[e.g.,][]{baumgardt_makino03, tiongco+16, bianchini+2016,
  bianchini+2018, webb_vesperini2017}, the diagnostics presented here
and those based on exotic populations as blue straggler stars
\citep[e.g.,][]{ferraro+18,ferraro+20} have the advantage to be more
easily measurable from observational data. In fact, they just require
high-resolution photometry down to $\sim 0.5$ magnitudes below the
main-sequence turnoff point and within a projected distance equal to
$0.5-1 \times r_h$ from the center.  This makes them suitable to
recognize highly evolved systems also in cases where the central
density cusp is not detectable from observations
\citep[e.g.,][]{Ferraro+23}, either because of too scarce statistics
in the construction of the density profile, or because the cluster is
not CC yet or is in a rebound stage during its post-CC evolution.
Because of their relatively simple determination, they also appear to
be promising tools for the investigation of the dynamical age of
stellar systems even beyond the Milky Way, out to distances where
individual stars in the upper main-sequence can be resolved.  A work
specifically devoted to the observational determination of the nCRD
parameters in a large sample of Galactic GCs is in preparation.


\section*{acknowledgments}  
This work is part of the project Cosmic-Lab at the Physics and
Astronomy Department "A. Righi" of the Bologna University
(http://www.cosmic-lab.eu/ Cosmic-Lab/Home.html).  AA acknowledges
support for this paper from project No. 2021/43/P/ST9/03167 co-funded
by the Polish National Science Center (NCN) and the European Union
Framework Programme for Research and Innovation Horizon 2020 under the
Marie Skłodowska-Curie grant agreement No. 945339. For the purpose of
Open Access, the authors have applied for a CC-BY public copyright
licence to any Author Accepted Manuscript (AAM) version arising from
this submission.  AH acknowledges support from grant
2021/41/B/ST9/01191 founded by Polish National Science Centre (NCN).
This research was supported in part by Lilly Endowment, Inc., through
its support for the Indiana University Pervasive Technology Institute.

\bibliography{ncrd_paper3}{}

\begin{thebibliography}{}
\expandafter\ifx\csname natexlab\endcsname\relax\def\natexlab#1{#1}\fi
\providecommand{\url}[1]{\href{#1}{#1}}
\providecommand{\dodoi}[1]{doi:~\href{http://doi.org/#1}{\nolinkurl{#1}}}
\providecommand{\doeprint}[1]{\href{http://ascl.net/#1}{\nolinkurl{http://ascl.net/#1}}}
\providecommand{\doarXiv}[1]{\href{https://arxiv.org/abs/#1}{\nolinkurl{https://arxiv.org/abs/#1}}}

\bibitem[{{Alessandrini} {et~al.}(2016){Alessandrini}, {Lanzoni}, {Ferraro}, {Miocchi}, \& {Vesperini}}]{alessandrini+16}
{Alessandrini}, E., {Lanzoni}, B., {Ferraro}, F.~R., {Miocchi}, P., \& {Vesperini}, E. 2016, \apj, 833, 252, \dodoi{10.3847/1538-4357/833/2/252}

\bibitem[{{Antonini} \& {Gieles}(2020)}]{antonini+2020}
{Antonini}, F., \& {Gieles}, M. 2020, \mnras, 492, 2936, \dodoi{10.1093/mnras/stz3584}

\bibitem[{{Arca Sedda} {et~al.}(2018){Arca Sedda}, {Askar}, \& {Giersz}}]{arcasedda+18}
{Arca Sedda}, M., {Askar}, A., \& {Giersz}, M. 2018, \mnras, 479, 4652, \dodoi{10.1093/mnras/sty1859}

\bibitem[{{Aros} \& {Vesperini}(2023)}]{aros+23}
{Aros}, F.~I., \& {Vesperini}, E. 2023, \mnras, 525, 3136, \dodoi{10.1093/mnras/stad2429}

\bibitem[{{Askar} {et~al.}(2018){Askar}, {Arca Sedda}, \& {Giersz}}]{askar+18}
{Askar}, A., {Arca Sedda}, M., \& {Giersz}, M. 2018, \mnras, 478, 1844, \dodoi{10.1093/mnras/sty1186}

\bibitem[{{Askar} {et~al.}(2017){Askar}, {Szkudlarek}, {Gondek-Rosi{\'n}ska}, {Giersz}, \& {Bulik}}]{askar+2017}
{Askar}, A., {Szkudlarek}, M., {Gondek-Rosi{\'n}ska}, D., {Giersz}, M., \& {Bulik}, T. 2017, \mnras, 464, L36, \dodoi{10.1093/mnrasl/slw177}

\bibitem[{{Baumgardt} \& {Makino}(2003)}]{baumgardt_makino03}
{Baumgardt}, H., \& {Makino}, J. 2003, \mnras, 340, 227, \dodoi{10.1046/j.1365-8711.2003.06286.x}

\bibitem[{{Baumgardt} {et~al.}(2005){Baumgardt}, {Makino}, \& {Hut}}]{baumgardt+05}
{Baumgardt}, H., {Makino}, J., \& {Hut}, P. 2005, \apj, 620, 238, \dodoi{10.1086/426893}

\bibitem[{{Belczynski} {et~al.}(2002){Belczynski}, {Kalogera}, \& {Bulik}}]{belczynski+02}
{Belczynski}, K., {Kalogera}, V., \& {Bulik}, T. 2002, \apj, 572, 407, \dodoi{10.1086/340304}

\bibitem[{{Bhat} {et~al.}(2022){Bhat}, {Lanzoni}, {Ferraro}, \& {Vesperini}}]{Bhat+22}
{Bhat}, B., {Lanzoni}, B., {Ferraro}, F.~R., \& {Vesperini}, E. 2022, \apj, 926, 118 (B22)

\bibitem[{{Bhat} {et~al.}(2023){Bhat}, {Lanzoni}, {Ferraro}, \& {Vesperini}}]{bhat+23}
---. 2023, \apj, 945, 164 (B23), \dodoi{10.3847/1538-4357/acb434}

\bibitem[{Bianchini {et~al.}(2016)Bianchini, van~de Ven, Norris, Schinnerer, \& Varri}]{bianchini+2016}
Bianchini, P., van~de Ven, G., Norris, M.~A., Schinnerer, E., \& Varri, A.~L. 2016, Monthly Notices of the Royal Astronomical Society, 458, 3644, \dodoi{10.1093/mnras/stw552}

\bibitem[{{Bianchini} {et~al.}(2018){Bianchini}, {Webb}, {Sills}, \& {Vesperini}}]{bianchini+2018}
{Bianchini}, P., {Webb}, J.~J., {Sills}, A., \& {Vesperini}, E. 2018, \mnras, 475, L96, \dodoi{10.1093/mnrasl/sly013}

\bibitem[{{Breen} \& {Heggie}(2013)}]{breen+13}
{Breen}, P.~G., \& {Heggie}, D.~C. 2013, \mnras, 432, 2779, \dodoi{10.1093/mnras/stt628}

\bibitem[{{Breen} {et~al.}(2017){Breen}, {Varri}, \& {Heggie}}]{breen+2017}
{Breen}, P.~G., {Varri}, A.~L., \& {Heggie}, D.~C. 2017, \mnras, 471, 2778, \dodoi{10.1093/mnras/stx1750}

\bibitem[{{Chernoff} \& {Djorgovski}(1989)}]{chernoff+89}
{Chernoff}, D.~F., \& {Djorgovski}, S. 1989, \apj, 339, 904, \dodoi{10.1086/167344}

\bibitem[{{Chernoff} \& {Weinberg}(1990)}]{chernoff+90}
{Chernoff}, D.~F., \& {Weinberg}, M.~D. 1990, \apj, 351, 121, \dodoi{10.1086/168451}

\bibitem[{{Djorgovski}(1993)}]{djorgovski93}
{Djorgovski}, S. 1993, in Astronomical Society of the Pacific Conference Series, Vol.~50, Structure and Dynamics of Globular Clusters, ed. S.~G. {Djorgovski} \& G.~{Meylan}, 373

\bibitem[{{Djorgovski} \& {King}(1984)}]{djorgovski+84}
{Djorgovski}, S., \& {King}, I.~R. 1984, \apjl, 277, L49, \dodoi{10.1086/184200}

\bibitem[{{Ferraro} {et~al.}(2020){Ferraro}, {Lanzoni}, \& {Dalessandro}}]{ferraro+20}
{Ferraro}, F.~R., {Lanzoni}, B., \& {Dalessandro}, E. 2020, Rendiconti Lincei. Scienze Fisiche e Naturali, 31, 19, \dodoi{10.1007/s12210-020-00873-2}

\bibitem[{{Ferraro} {et~al.}(2019){Ferraro}, {Lanzoni}, {Dalessandro}, {Cadelano}, {Raso}, {Mucciarelli}, {Beccari}, \& {Pallanca}}]{ferraro+19}
{Ferraro}, F.~R., {Lanzoni}, B., {Dalessandro}, E., {et~al.} 2019, Nature Astronomy, 3, 1149, \dodoi{10.1038/s41550-019-0865-1}

\bibitem[{{Ferraro} {et~al.}(2023){Ferraro}, {Lanzoni}, {Vesperini}, {Cadelano}, {Deras}, \& {Pallanca}}]{Ferraro+23}
{Ferraro}, F.~R., {Lanzoni}, B., {Vesperini}, E., {et~al.} 2023, \apj, 950, 145, \dodoi{10.3847/1538-4357/accd5c}

\bibitem[{{Ferraro} {et~al.}(2012){Ferraro}, {Lanzoni}, {Dalessandro}, {Beccari}, {Pasquato}, {Miocchi}, {Rood}, {Sigurdsson}, {Sills}, {Vesperini}, {Mapelli}, {Contreras}, {Sanna}, \& {Mucciarelli}}]{ferraro+12}
{Ferraro}, F.~R., {Lanzoni}, B., {Dalessandro}, E., {et~al.} 2012, \nat, 492, 393, \dodoi{10.1038/nature11686}

\bibitem[{{Ferraro} {et~al.}(2018){Ferraro}, {Lanzoni}, {Raso}, {Nardiello}, {Dalessandro}, {Vesperini}, {Piotto}, {Pallanca}, {Beccari}, {Bellini}, {Libralato}, {Anderson}, {Aparicio}, {Bedin}, {Cassisi}, {Milone}, {Ortolani}, {Renzini}, {Salaris}, \& {van der Marel}}]{ferraro+18}
{Ferraro}, F.~R., {Lanzoni}, B., {Raso}, S., {et~al.} 2018, \apj, 860, 36, \dodoi{10.3847/1538-4357/aac01c}

\bibitem[{{Gieles} {et~al.}(2021){Gieles}, {Erkal}, {Antonini}, {Balbinot}, \& {Pe{\~n}arrubia}}]{gieles+2021}
{Gieles}, M., {Erkal}, D., {Antonini}, F., {Balbinot}, E., \& {Pe{\~n}arrubia}, J. 2021, Nature Astronomy, 5, 957, \dodoi{10.1038/s41550-021-01392-2}

\bibitem[{Giersz {et~al.}(2013)Giersz, Heggie, Hurley, \& Hypki}]{Mocca_giersz}
Giersz, M., Heggie, D.~C., Hurley, J.~R., \& Hypki, A. 2013, Monthly Notices of the Royal Astronomical Society, 431, 2184, \dodoi{10.1093/mnras/stt307}

\bibitem[{{Giersz} {et~al.}(2015){Giersz}, {Leigh}, {Hypki}, {L{\"u}tzgendorf}, \& {Askar}}]{giersz+15}
{Giersz}, M., {Leigh}, N., {Hypki}, A., {L{\"u}tzgendorf}, N., \& {Askar}, A. 2015, \mnras, 454, 3150, \dodoi{10.1093/mnras/stv2162}

\bibitem[{{Gill} {et~al.}(2008){Gill}, {Trenti}, {Miller}, {van der Marel}, {Hamilton}, \& {Stiavelli}}]{Gill+08}
{Gill}, M., {Trenti}, M., {Miller}, M.~C., {et~al.} 2008, \apj, 686, 303, \dodoi{10.1086/591269}

\bibitem[{{Harris}(1996)}]{harris96}
{Harris}, W.~E. 1996, \aj, 112, 1487, \dodoi{10.1086/118116}

\bibitem[{{Heggie} \& {Hut}(2003)}]{heggiehut2003}
{Heggie}, D., \& {Hut}, P. 2003, {The Gravitational Million-Body Problem: A Multidisciplinary Approach to Star Cluster Dynamics}

\bibitem[{Hobbs {et~al.}(2005)Hobbs, Lorimer, Lyne, \& Kramer}]{Hobbs_05}
Hobbs, G., Lorimer, D.~R., Lyne, A.~G., \& Kramer, M. 2005, Monthly Notices of the Royal Astronomical Society, 360, 974, \dodoi{10.1111/j.1365-2966.2005.09087.x}

\bibitem[{{Hong} {et~al.}(2018){Hong}, {Vesperini}, {Askar}, {Giersz}, {Szkudlarek}, \& {Bulik}}]{hong+2018}
{Hong}, J., {Vesperini}, E., {Askar}, A., {et~al.} 2018, \mnras, 480, 5645, \dodoi{10.1093/mnras/sty2211}

\bibitem[{{Hong} {et~al.}(2017){Hong}, {Vesperini}, {Belloni}, \& {Giersz}}]{hong+2017}
{Hong}, J., {Vesperini}, E., {Belloni}, D., \& {Giersz}, M. 2017, \mnras, 464, 2511, \dodoi{10.1093/mnras/stw2595}

\bibitem[{{Humphreys} \& {Davidson}(1994)}]{Humphreys+94}
{Humphreys}, R.~M., \& {Davidson}, K. 1994, \pasp, 106, 1025, \dodoi{10.1086/133478}

\bibitem[{{Hurley} {et~al.}(2000){Hurley}, {Pols}, \& {Tout}}]{hurley+00}
{Hurley}, J.~R., {Pols}, O.~R., \& {Tout}, C.~A. 2000, \mnras, 315, 543, \dodoi{10.1046/j.1365-8711.2000.03426.x}

\bibitem[{{Hurley} {et~al.}(2002){Hurley}, {Tout}, \& {Pols}}]{hurley+02}
{Hurley}, J.~R., {Tout}, C.~A., \& {Pols}, O.~R. 2002, \mnras, 329, 897, \dodoi{10.1046/j.1365-8711.2002.05038.x}

\bibitem[{{Hypki} \& {Giersz}(2013)}]{Hypki_Giersz_2013}
{Hypki}, A., \& {Giersz}, M. 2013, \mnras, 429, 1221, \dodoi{10.1093/mnras/sts415}

\bibitem[{King(1966)}]{king66}
King, I.~R. 1966, The Astronomical Journal, 71, 64

\bibitem[{{Kremer} {et~al.}(2021){Kremer}, {Rui}, {Weatherford}, {Chatterjee}, {Fragione}, {Rasio}, {Rodriguez}, \& {Ye}}]{kremer+21}
{Kremer}, K., {Rui}, N.~Z., {Weatherford}, N.~C., {et~al.} 2021, \apj, 917, 28, \dodoi{10.3847/1538-4357/ac06d4}

\bibitem[{{Kremer} {et~al.}(2018){Kremer}, {Ye}, {Chatterjee}, {Rodriguez}, \& {Rasio}}]{kremer+2018}
{Kremer}, K., {Ye}, C.~S., {Chatterjee}, S., {Rodriguez}, C.~L., \& {Rasio}, F.~A. 2018, \apjl, 855, L15, \dodoi{10.3847/2041-8213/aab26c}

\bibitem[{{Kremer} {et~al.}(2020){Kremer}, {Ye}, {Rui}, {Weatherford}, {Chatterjee}, {Fragione}, {Rodriguez}, {Spera}, \& {Rasio}}]{kremer+20}
{Kremer}, K., {Ye}, C.~S., {Rui}, N.~Z., {et~al.} 2020, \apjs, 247, 48, \dodoi{10.3847/1538-4365/ab7919}

\bibitem[{{Kroupa}(1995)}]{kroupa95}
{Kroupa}, P. 1995, \mnras, 277, 1491, \dodoi{10.1093/mnras/277.4.1491}

\bibitem[{{Kroupa}(2001)}]{IMF}
---. 2001, \mnras, 322, 231

\bibitem[{{Kroupa} {et~al.}(2013){Kroupa}, {Weidner}, {Pflamm-Altenburg}, {Thies}, {Dabringhausen}, {Marks}, \& {Maschberger}}]{kroupa+2013}
{Kroupa}, P., {Weidner}, C., {Pflamm-Altenburg}, J., {et~al.} 2013, in Planets, Stars and Stellar Systems. Volume 5: Galactic Structure and Stellar Populations, ed. T.~D. {Oswalt} \& G.~{Gilmore}, Vol.~5, 115, \dodoi{10.1007/978-94-007-5612-0_4}

\bibitem[{{Lanzoni} {et~al.}(2016){Lanzoni}, {Ferraro}, {Alessandrini}, {Dalessandro}, {Vesperini}, \& {Raso}}]{lanzoni+16}
{Lanzoni}, B., {Ferraro}, F.~R., {Alessandrini}, E., {et~al.} 2016, \apjl, 833, L29, \dodoi{10.3847/2041-8213/833/2/L29}

\bibitem[{{Lanzoni} {et~al.}(2019){Lanzoni}, {Ferraro}, {Dalessandro}, {Cadelano}, {Pallanca}, {Raso}, {Mucciarelli}, {Beccari}, \& {Focardi}}]{Lanzoni+19}
{Lanzoni}, B., {Ferraro}, F.~R., {Dalessandro}, E., {et~al.} 2019, \apj, 887, 176, \dodoi{10.3847/1538-4357/ab54c2}

\bibitem[{{Leveque} {et~al.}(2021){Leveque}, {Giersz}, \& {Paolillo}}]{leveque+21}
{Leveque}, A., {Giersz}, M., \& {Paolillo}, M. 2021, \mnras, 501, 5212, \dodoi{10.1093/mnras/staa4027}

\bibitem[{{Mackey} {et~al.}(2007){Mackey}, {Wilkinson}, {Davies}, \& {Gilmore}}]{mackey+07}
{Mackey}, A.~D., {Wilkinson}, M.~I., {Davies}, M.~B., \& {Gilmore}, G.~F. 2007, \mnras, 379, L40, \dodoi{10.1111/j.1745-3933.2007.00330.x}

\bibitem[{{Makino}(1996)}]{makino+1996}
{Makino}, J. 1996, \apj, 471, 796, \dodoi{10.1086/178007}

\bibitem[{{Miocchi}(2007)}]{miocchi07}
{Miocchi}, P. 2007, \mnras, 381, 103, \dodoi{10.1111/j.1365-2966.2007.12165.x}

\bibitem[{Miocchi {et~al.}(2013)Miocchi, Lanzoni, Ferraro, Dalessandro, Vesperini, Pasquato, Beccari, Pallanca, \& Sanna}]{miocchi+13}
Miocchi, P., Lanzoni, B., Ferraro, F.~R., {et~al.} 2013, The Astrophysical Journal, 774, 151, \dodoi{10.1088/0004-637x/774/2/151}

\bibitem[{{Nieuwenhuijzen} \& {de Jager}(1990)}]{nieuwenhuijzen+90}
{Nieuwenhuijzen}, H., \& {de Jager}, C. 1990, \aap, 231, 134

\bibitem[{{Pavl{\'\i}k} \& {Vesperini}(2022)}]{pavlik+2022}
{Pavl{\'\i}k}, V., \& {Vesperini}, E. 2022, \mnras, 509, 3815, \dodoi{10.1093/mnras/stab3157}

\bibitem[{{Peuten} {et~al.}(2016){Peuten}, {Zocchi}, {Gieles}, {Gualandris}, \& {H{\'e}nault-Brunet}}]{peuten+2016}
{Peuten}, M., {Zocchi}, A., {Gieles}, M., {Gualandris}, A., \& {H{\'e}nault-Brunet}, V. 2016, \mnras, 462, 2333, \dodoi{10.1093/mnras/stw1726}

\bibitem[{{Rodriguez} {et~al.}(2016){Rodriguez}, {Chatterjee}, \& {Rasio}}]{rodriguez+2016}
{Rodriguez}, C.~L., {Chatterjee}, S., \& {Rasio}, F.~A. 2016, \prd, 93, 084029, \dodoi{10.1103/PhysRevD.93.084029}

\bibitem[{{Rodriguez} {et~al.}(2015){Rodriguez}, {Morscher}, {Pattabiraman}, {Chatterjee}, {Haster}, \& {Rasio}}]{rodriguez+2015}
{Rodriguez}, C.~L., {Morscher}, M., {Pattabiraman}, B., {et~al.} 2015, \prl, 115, 051101, \dodoi{10.1103/PhysRevLett.115.051101}

\bibitem[{{Tiongco} {et~al.}(2016){Tiongco}, {Vesperini}, \& {Varri}}]{tiongco+16}
{Tiongco}, M.~A., {Vesperini}, E., \& {Varri}, A.~L. 2016, \mnras, 455, 3693, \dodoi{10.1093/mnras/stv2574}

\bibitem[{{Trager} {et~al.}(1993){Trager}, {Djorgovski}, \& {King}}]{trager+93}
{Trager}, S.~C., {Djorgovski}, S., \& {King}, I.~R. 1993, in Astronomical Society of the Pacific Conference Series, Vol.~50, Structure and Dynamics of Globular Clusters, ed. S.~G. {Djorgovski} \& G.~{Meylan}, 347

\bibitem[{{van den Bosch} {et~al.}(2006){van den Bosch}, {de Zeeuw}, {Gebhardt}, {Noyola}, \& {van de Ven}}]{vandenbosch+06}
{van den Bosch}, R., {de Zeeuw}, T., {Gebhardt}, K., {Noyola}, E., \& {van de Ven}, G. 2006, \apj, 641, 852, \dodoi{10.1086/500644}

\bibitem[{{Vesperini} \& {Trenti}(2010)}]{Vesperini+10}
{Vesperini}, E., \& {Trenti}, M. 2010, \apjl, 720, L179, \dodoi{10.1088/2041-8205/720/2/L179}

\bibitem[{{Vitral} {et~al.}(2023){Vitral}, {Libralato}, {Kremer}, {Mamon}, {Bellini}, {Bedin}, \& {Anderson}}]{vitral+23}
{Vitral}, E., {Libralato}, M., {Kremer}, K., {et~al.} 2023, \mnras, 522, 5740, \dodoi{10.1093/mnras/stad1068}

\bibitem[{{Weatherford} {et~al.}(2018){Weatherford}, {Chatterjee}, {Rodriguez}, \& {Rasio}}]{weatherford+18}
{Weatherford}, N.~C., {Chatterjee}, S., {Rodriguez}, C.~L., \& {Rasio}, F.~A. 2018, \apj, 864, 13, \dodoi{10.3847/1538-4357/aad63d}

\bibitem[{{Webb} \& {Vesperini}(2017)}]{webb_vesperini2017}
{Webb}, J.~J., \& {Vesperini}, E. 2017, \mnras, 464, 1977, \dodoi{10.1093/mnras/stw2513}

\bibitem[{{Ye} {et~al.}(2020){Ye}, {Fong}, {Kremer}, {Rodriguez}, {Chatterjee}, {Fragione}, \& {Rasio}}]{ye+20}
{Ye}, C.~S., {Fong}, W.-f., {Kremer}, K., {et~al.} 2020, \apjl, 888, L10, \dodoi{10.3847/2041-8213/ab5dc5}

\end{thebibliography}
\bibliographystyle{aasjournal}

\end{document}